\documentclass{raa}          

\usepackage{graphicx,times,epsf}             %for PS/EPS graphics inclusion, new
\usepackage{amsmath}
\usepackage{makecell}
\usepackage[utf8]{inputenc}
\usepackage{multirow}
\newcommand{\Msol}{M$_{\odot}$}
\newcommand{\Msold}{M$_{\odot}$\,yr$^{-1}$}

\begin{document}
   \title{Morphology and kinematics of the gas envelope of the Mira binary W Aquilae
\thanks{This paper makes use of the following ALMA data: \mbox{ADS/JAO.ALMA$\#$$<$2012.1.00524.S$>$}. 
ALMA is a partnership of ESO (representing its member states), NSF (USA) and NINS (Japan), 
together with NRC (Canada) and NSC and ASIAA (Taiwan), and KASI (Republic of Korea), in cooperation with the Republic of Chile. 
The Joint ALMA Observatory is operated by ESO, AUI/NRAO and NAOJ. The data are retrieved 
from the JVO portal (http://jvo.nao.ac.jp/portal) operated by the NAOJ}}
%   \subtitle{I. Place Your Subtitle Here}
   \volnopage{Vol.0 (200x) No.0, 000--000}      %%preserved for Editor. DOn't remove!
   \setcounter{page}{1}          %%starting page, preserved for Editor. DOn't remove!  
   \author{D. T. Hoai\inst{}
          \and
          P. T. Nhung\inst{}
          \and
          P. N. Diep\inst{}
          \and
          N. T. Phuong\inst{}
          \and
          P. Tuan-Anh\inst{}
          \and
          N. T. Thao\inst{}
          \and
          P. Darriulat\inst{}
     }     
   \institute{Department of Astrophysics, Vietnam National Satellite Center, VAST, 18 Hoang Quoc Viet, Cau Giay, Hanoi, Vietnam; {\it dthoai@vnsc.org.vn}}
   \date{}
   \abstract {We analyse ALMA observations of the {$^{12}$CO(3-2)} emission of the circumstellar envelope 
(CSE) of the Mira variable binary star W Aql. These provide, for the first time, spatially resolved Doppler 
velocity spectra of the CSE up to angular distances to the central star of $\sim5''$ (meaning some 2000 AU). 
The exploratory nature of the observations (only five minutes in each of two different configurations) does 
not allow for a detailed modelling of the properties of the CSE but provides important qualitative information
 on its morphology and kinematics. Emission is found to be enhanced along an axis moving from east/west to 
north-east/south-west when the angular distance from the central star projected on the plane of the sky 
increases from zero to four arcseconds. In parallel, the Doppler velocity distribution displays asymmetry 
along an axis moving from east/west to north-west/south-east. The results are discussed in the context of 
earlier observations, in particular of the dust morphology.
   }
   \authorrunning{D. T. Hoai et al.}            %author_head in even pages
   \titlerunning{}  % title_head in odd pages
   \maketitle
%% The author head (on even pages) and the title head (on odd pages) will be
%% automatically extracted from \author{} and \title{}. Whenever the title is too long,
%% you will be asked to supply a shorter one by inserting either \authorrunning{} or
%% \titlerunning{} before \maketitle. Anyway, you can specify your own heads.
%%
%%
%% Note: In the following text body of your manuscript, please note several differences from
%%       other major journals:
%% (1) \subsection{Please Capitalize the First Letter of Each Notional Word in Subsection Title}
%% (2) Please Capitalize the First Letter of Each Notional Word in all tables' captions

%
%________________________________________________ sections below

\section{Introduction}
W Aql is a Mira variable star of spectral type S6,6, namely in the transition between being oxygen and carbon dominated,
 with a period of 490 days, a mean luminosity of $\sim$7000 solar luminosities and an effective temperature of $\sim$2000 K 
(\cite{SP1998}, \cite{Kuk1971}, \cite{FW2000}, \cite{Alf2012}, \cite{Ram2009}, \cite{Dan2014}).

\begin{figure}[!ht]
\begin{center}
%trim option's parameter order: left bottom right top
\includegraphics[scale=0.28]{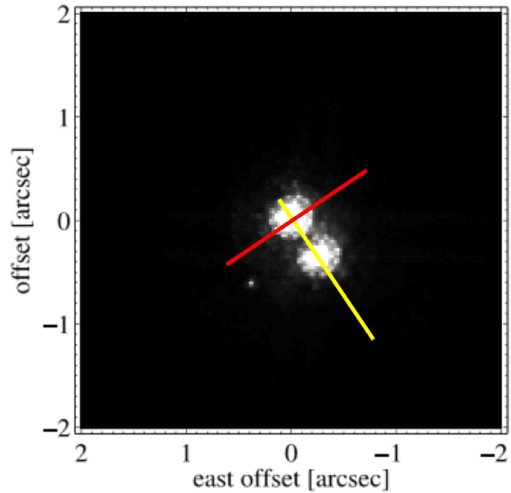}
\caption{High-resolution HST B-band image of W Aql. The binary system is clearly resolved and the AGB star is centred. 
North is up and east is left. The yellow line, joining the central star to its companion, is at 56$^\circ$ clockwise from 
east/west, the red line at 34$^\circ$ counter-clockwise.}\label{Fig1}
\end{center}
\end{figure}

It is one of the few AGB stars for which direct evidence for the presence of a companion has been obtained (\cite{Ram2011}). 
A high-resolution Hubble Space Telescope\footnote{Based on observations made with the NASA/ESA Hubble Space Telescope,
 obtained at the data archive at the Space Telescope Science Institute. STScI is operated by the Association 
of Universities for Research in Astronomy, Inc. under NASA contract NAS 5-26555.} B-band image taken in 2004 at 435 nm 
(Figure \ref{Fig1}) displays a clearly resolved binary system, the companion being located in the south-west direction
 at about 56$^\circ$ from west and at a projected distance of \mbox{$\sim0.47''$} from the central star. The companion
 is a main sequence star with an effective temperature of $\sim$6000 K and a mass of $\sim$1 \Msol 
(\cite{Dan2015}, \cite{May2013}). The distance from Earth is poorly known, with values between 230 pc and 610 pc 
being quoted in the literature (Guandalini \& Busso 2008, Wallerstein et al. 2011, Knapp et al. 1998, Loup et al. 1993, 
Keenan \& Boeshaar 1980, Danilovich et al. 2014, Ramstedt et al. 2009, 2011, Ramstedt \& Oloffson 2014), indirectly 
obtained from the luminosity-period relation (Whitelock et al. 1994, 2008). Retaining a distance of $\sim$400 pc, the
 projected distance between the central star and its companion is of the order of $\sim$200 AU. The real distance may
 be much larger as nothing is known of the inclination of the orbital plane on the sky. As two HST images separated by
 11 years do not reveal any displacement between star and companion, the orbital period is estimated to exceed $\sim$1000 years 
(\cite{May2013}, \cite{Dan2015}).

The star has been extensively observed in the visible, the infrared, down to far infrared, and radio. Ramstedt et al. (2011) 
have recently detected polarized light in the visible, namely essentially light from the star scattered from the dust in
 its CSE. As polarization is strongly peaked about 90$^\circ$ scattering, they observe the circumstellar dust in a slice 
containing the central star and parallel to the sky plane. Their observations reveal a clear enhancement of the emission
 in the south-west direction up to an angular distance of some $10''$ from the central star. Close to the star, up to a
 distance of \mbox{$\sim1''$}, the emission is clearly elongated around the north-east/south-west direction. While this 
direction corresponds to the line joining the central star to its companion in Figure \ref{Fig1}, a causal relation between
 the two directions is far from obvious as the inclination of the orbit on the sky plane is unknown. The dust mass of the 
south-west feature is estimated to be (assuming optically thin dust-scattering) of the order of 10$^{-6}$ \Msol.

%%%%%%%%%%%%%%%   fig 1 %%%%%%%%%%%%%%%%%%%%%

Radio single dish observations in the millimetre/sub-millimetre range include several molecular lines (\cite{Nym1992}, 
\cite{Kna1998}, \cite{Ram2009}, \cite{Wall2011} and \cite{RO2014}). The associated line profiles cover typically a 
$\pm$20 km\,s$^{-1}$ Doppler velocity interval, giving evidence for significant winds. Danilovich et al. (2015) have 
recently modelled the line profiles of $^{12}$CO, $^{13}$CO, SiO, H$_2$O, HCN and NH$_3$, and obtained a mass loss 
rate of 4.0 10$^{-6}$ \Msold, of the same order of magnitude as previous estimates that range between 
2.2 10$^{-6}$ and 9.4 10$^{-6}$ \Msold (\cite{Kna1998}, \cite{Ram2009}, \cite{De2010}). The profiles are 
approximately symmetric with an occasional slight excess emission on the blue-shifted side, suggesting an asymmetric 
outflow. Continuum emission has also been observed and the dust properties have been studied up to large distances from
 the central star (\cite{May2013}). The only interferometer observations that have been made are from the Berkeley
 Infrared Spatial Interferometer (ISI) at \mbox{11 $\mu$m} and the Plateau de Bure Interferometer on CO molecular lines.
 The former use crude interferometry from three aligned antennae and find an excess of emission in the eastern half-plane 
for small distances from the central star, reaching to $0.5''$ (\cite{Tat2006}); the latter are part of the COSAS programme
 (\cite{Cas2010}) but have not been published yet.

Recently, ALMA observations of the CO(3-2) emission of the circumstellar envelope have been made publicly available
 (ALMA 0100334[3/5]). The present work presents an analysis of these data, which provide for the first time spatially
 resolved Doppler velocity spectra.

\section{Observations, data reduction and data selection}
\label{sec:Observations}
Two sets of observations were made on March 6$^{\text{th}}$ and 20$^{\text{th}}$, 2014, each of five minutes duration, 
one with the normal array (ALMA) and the other with the compact array (ACA) with configurations giving beams of 
respective angular dimensions of $1.1''$$\times$$1.0''$ and $4.9''$$\times$$2.6''$ and position angles of $-77.6^\circ$
 and $75.6^\circ$. The line Doppler velocity spectra, are available in each case in bins of \mbox{$2$ km\,s$^{-1}$} 
between \mbox{$-55.5$ km\,s$^{-1}$} and \mbox{$4.5$ km\,s$^{-1}$}. In what follows, we define the middle of the range, 
\mbox{$-25.5$ km\,s$^{-1}$}, as origin of velocity coordinates. Also available are continuum flux densities averaged 
over a 15.8 GHz interval. The data have been reduced by the ALMA staff to clean maps using pixel sizes of $0.1''\times0.1''$
 (ALMA) and $0.8''\times0.8''$ (ACA) respectively. We use orthonormal coordinates with $x$ along the line of sight 
away from Earth, $y$ pointing east and $z$ pointing north.

%%%%%%% fig 2 %%%%%%%%%%%%%
\begin{figure}[!ht]
\begin{center}
\includegraphics[scale=0.36]{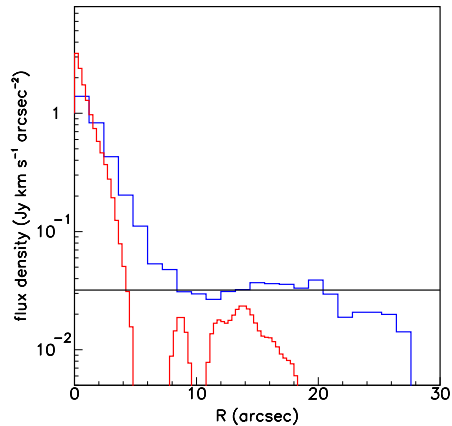}
\caption{Distributions of the line flux integrated over Doppler velocities and averaged over position angles $\varphi$ 
as a function of $R=\sqrt{y^2+z^2}$ for ALMA data (red) and ACA data (blue). The black line is for a flux of \mbox{32 mJy km\,s$^{-1}$\,arcsec$^{-2}$}.}\label{Fig2}
\end{center}
\end{figure}
%%%%%%% fig 3 %%%%%%%%%%%%%
\begin{figure*}[!ht]
\begin{center}
\includegraphics[height=4cm]{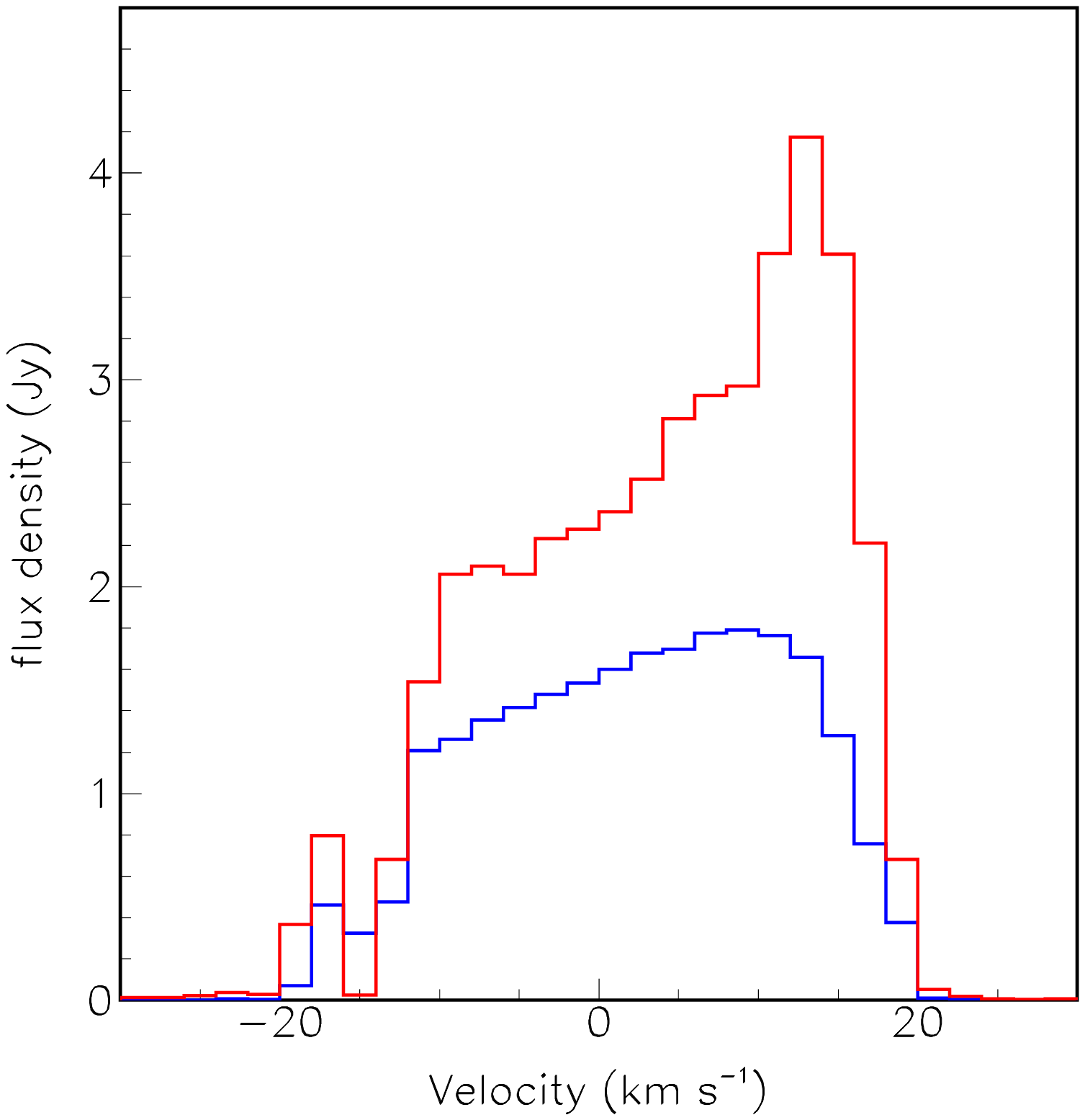}
\includegraphics[height=4 cm]{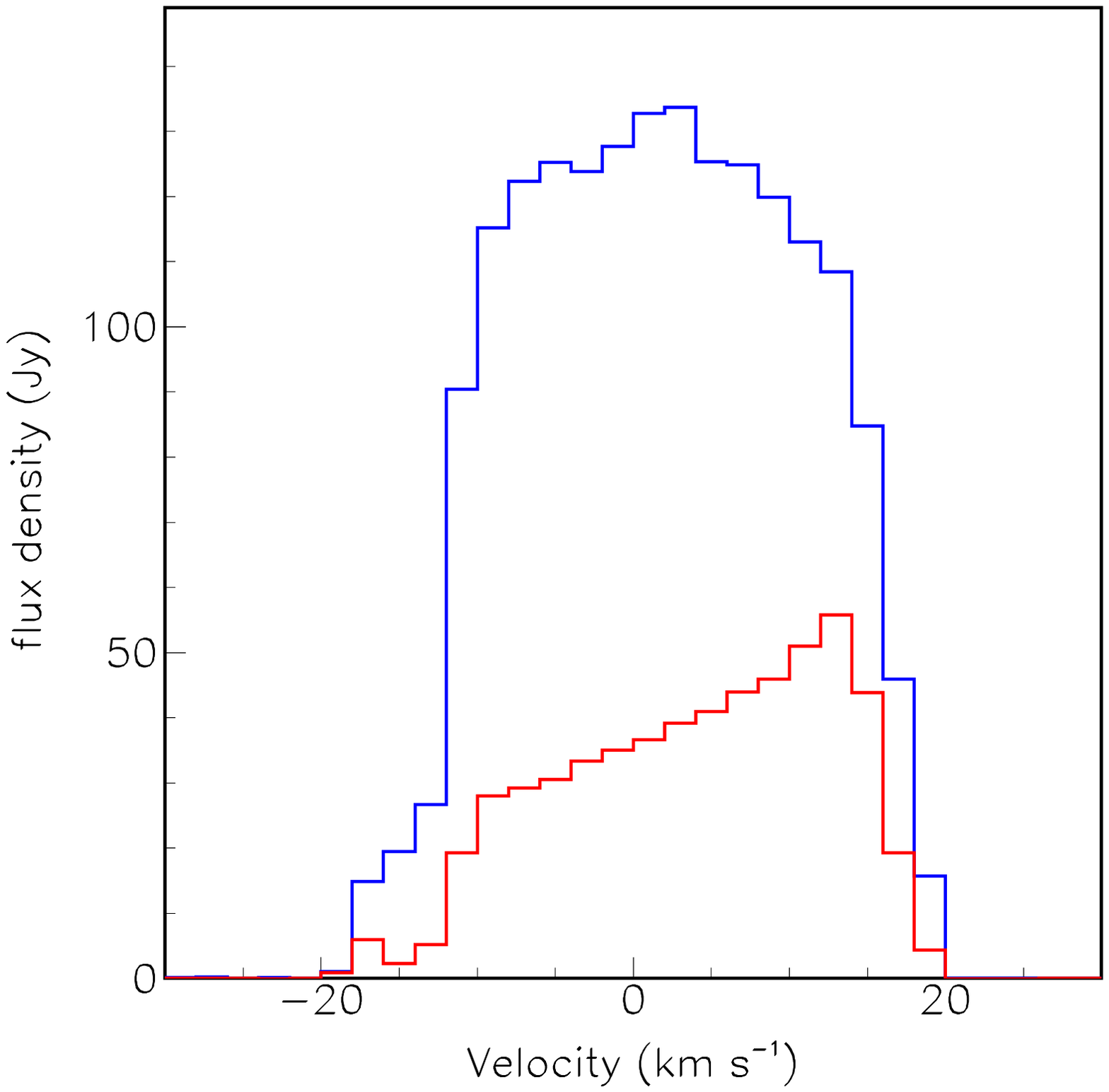}
\includegraphics[height=4 cm,trim = 0cm -1.cm 0.cm 0.cm, clip]{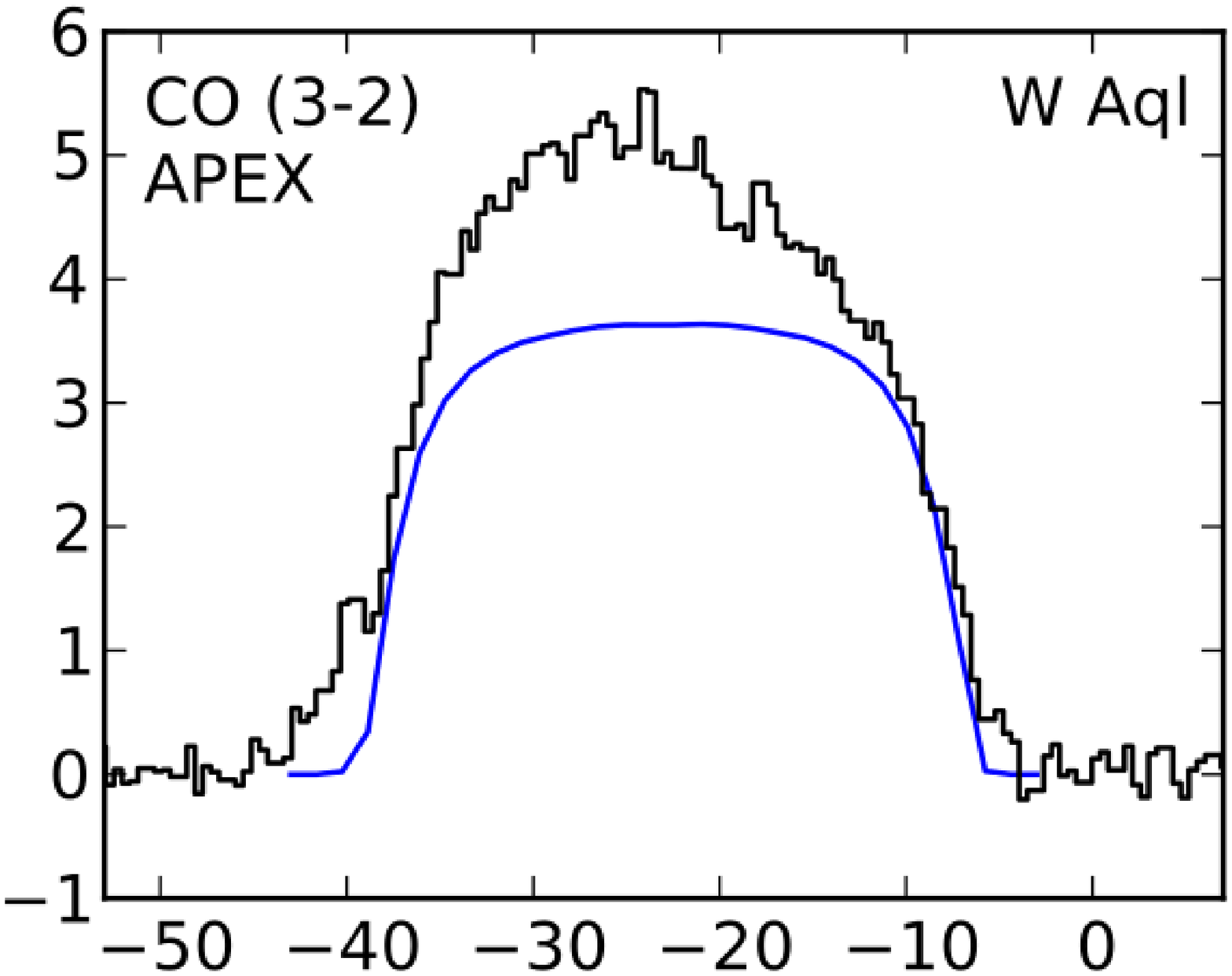}
\caption{Dependence of the line flux density on Doppler velocity averaged over the central $1.6''\times1.6''$ (left panel) 
and integrated over a Gaussian beam of $18''$ FWHM (central panel) for ALMA (red) and ACA (blue) data. The right panel 
shows the profile measured by Ramstedt et al. (2009) together with the model fit of Danilovich et al. (2015).}\label{Fig3}
\end{center}
\end{figure*}

Figure \ref{Fig2} shows the distributions of the line flux integrated over Doppler velocities and averaged over position angle
$\varphi=\tan^{-1} (\frac{z}{y})+\pi$ ($=0$ west and positive clockwise) as a function of $R=\sqrt{y^2 +z^2}$, the projected 
angular distance from the star. It gives evidence at \mbox{$\sim4''$} for a short spacing cut-off in the ALMA data, namely the
 flux filtered out by the interferometer, and for extended emission in the ACA data extending up to at least $22''$ at a typical 
level of \mbox{$\sim32$ mJy km\,s$^{-1}$\,arcsec$^{-2}$} (a constant level in $R$ means a $1/r$ dependence in space). 
Accordingly, in what follows, we restrict the analysis of the ALMA data to angular distances from the central star not exceeding
 $4''$ (meaning $\sim$ 1600 AU in projected distance).

%%%%%%%%%%%%%%%%% Figure 4 %%%%%%%%%%%%%%
\begin{figure*}[!ht]
\begin{center}
\includegraphics[scale=0.34]{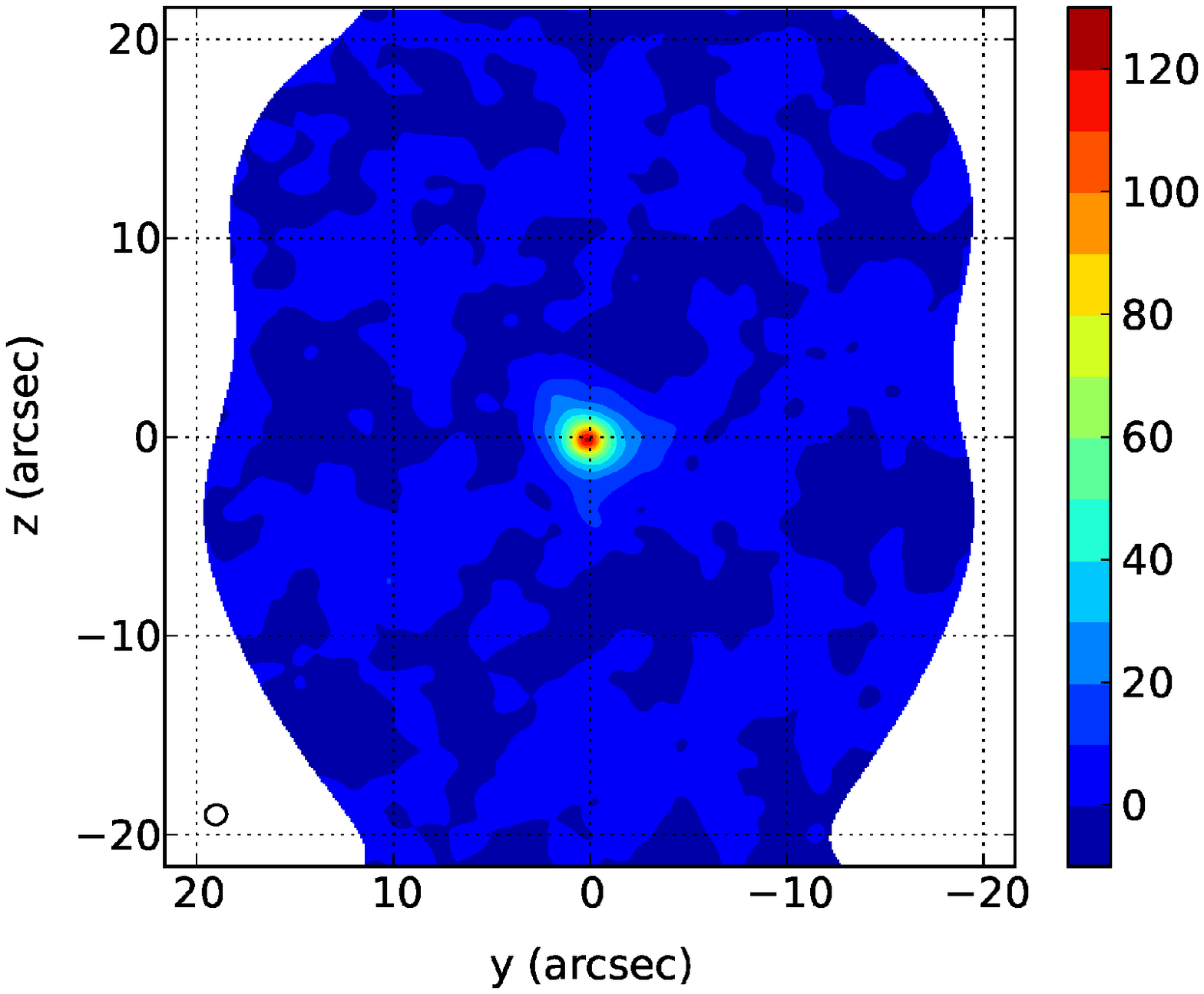}
\includegraphics[scale=0.34]{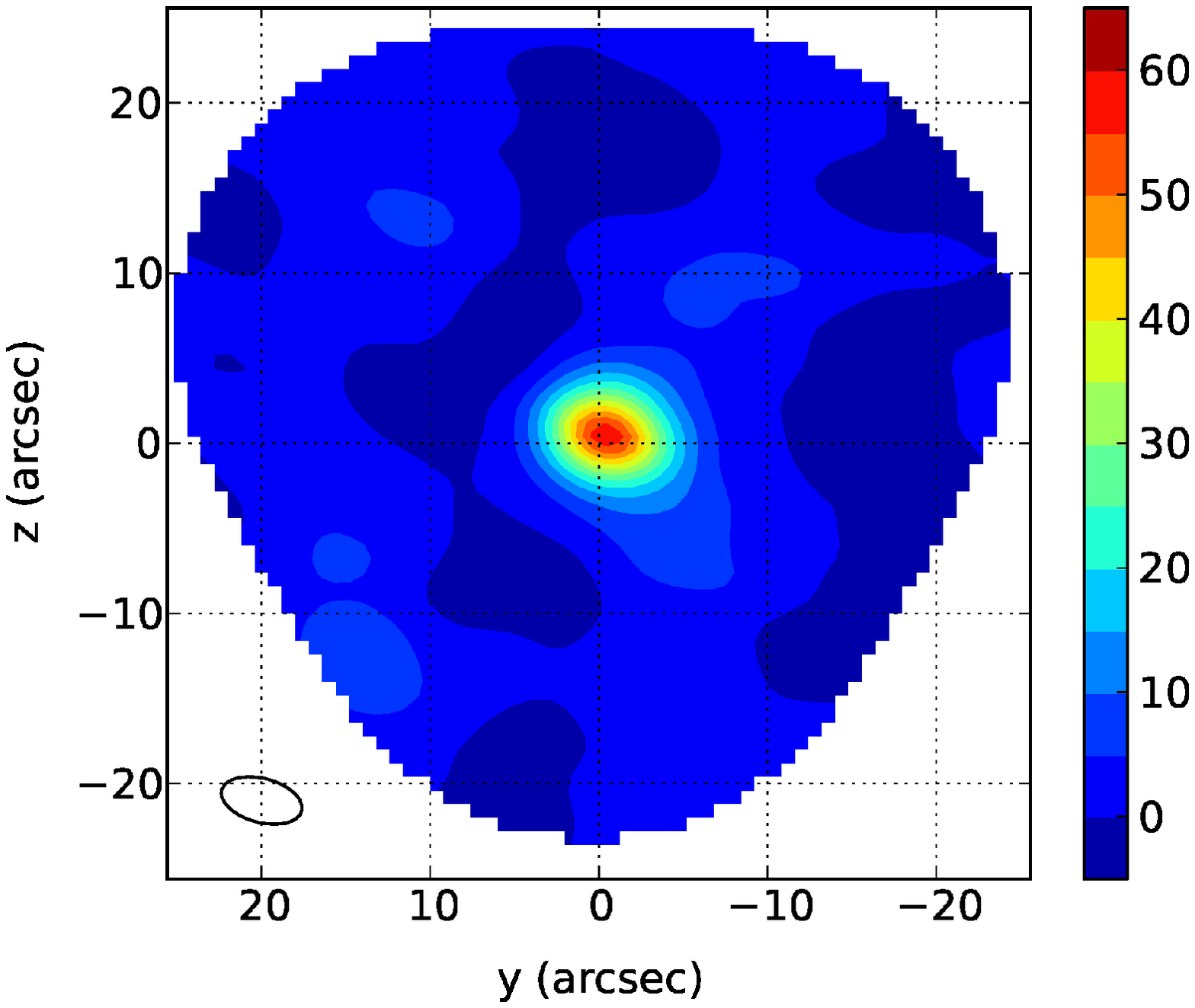}
\includegraphics[scale=0.34]{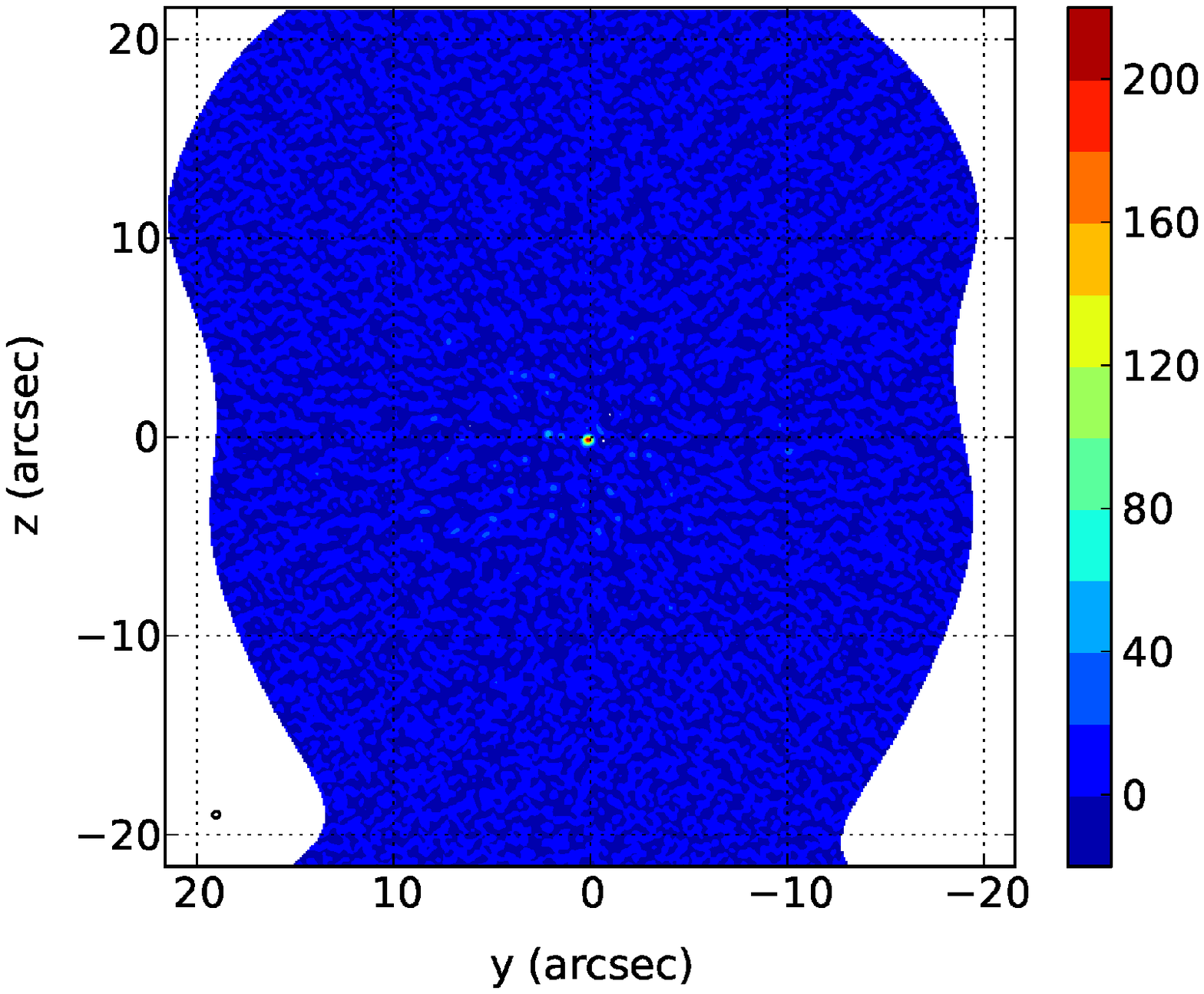}
\includegraphics[scale=0.34]{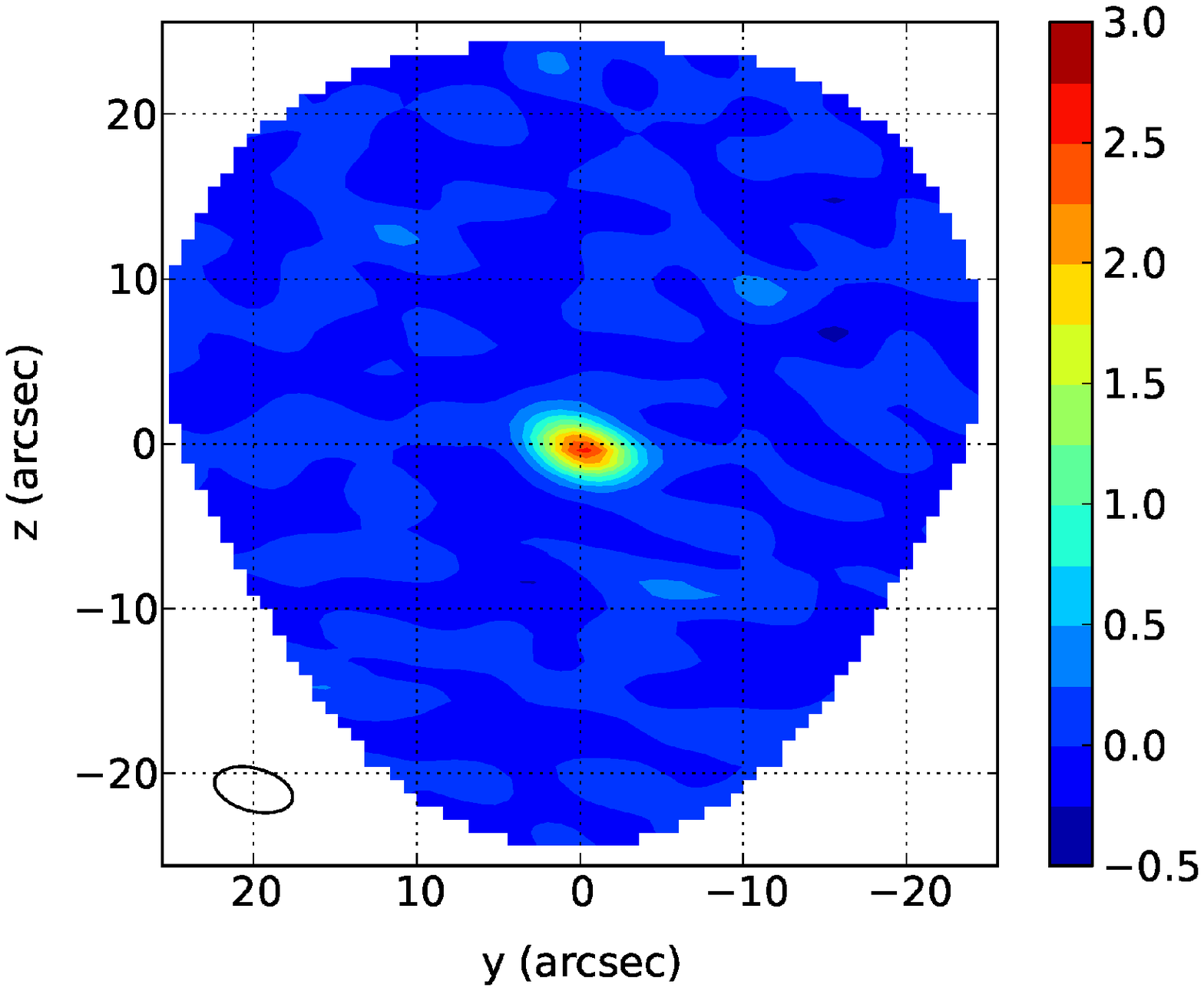}
\caption{Sky maps of the detected fluxes for the line (upper panels) and the continuum (lower panels) and for ALMA data (left panels) 
and ACA data (right panels) separately. The beam is shown in the lower left corner. The color scales are in units of Jy km\,s$^{-1}$\,arcsec$^{-2}$
 for the line and mJy\,arcsec$^{-2}$ for the continuum.}\label{Fig4}
\end{center}
\end{figure*}

Figure \ref{Fig3} displays the dependence of the line flux density on Doppler velocity averaged over the central $1.6''\times1.6''$
 (left panel) and over a Gaussian circular beam having a FWHM of $18''$ (central panel). The latter corresponds to the APEX single 
dish observation of Ramstedt et al. (2009) which is also displayed in the figure (right panel). Significant differences between 
the ALMA and ACA data are the effect of the important short spacing problem in the former. In particular, the data integrated 
over the large Gaussian beam show directly that more than half of the total flux is contributed by the large distance environment
 of the star, at distances exceeding $4''$. While the line profile observed by ACA is in good agreement with that observed in the
 single dish APEX data, evidence for an important slope in the central region is obtained, the red-shifted side being enhanced with 
respect to the blue-shifted side by $\sim2$\% per km\,s$^{-1}$. The slope is also apparent in the ACA data when they are restricted 
to the central region (here a square of $1.6''\times1.6''$, significantly smaller than the beam size).

%%%%%%%%%%%%%%%%%%%Figure 5%%%%%%%%%%%%%%%%%%%%%%%
\begin{figure}[!ht]
\begin{center}
\includegraphics[scale=0.36]{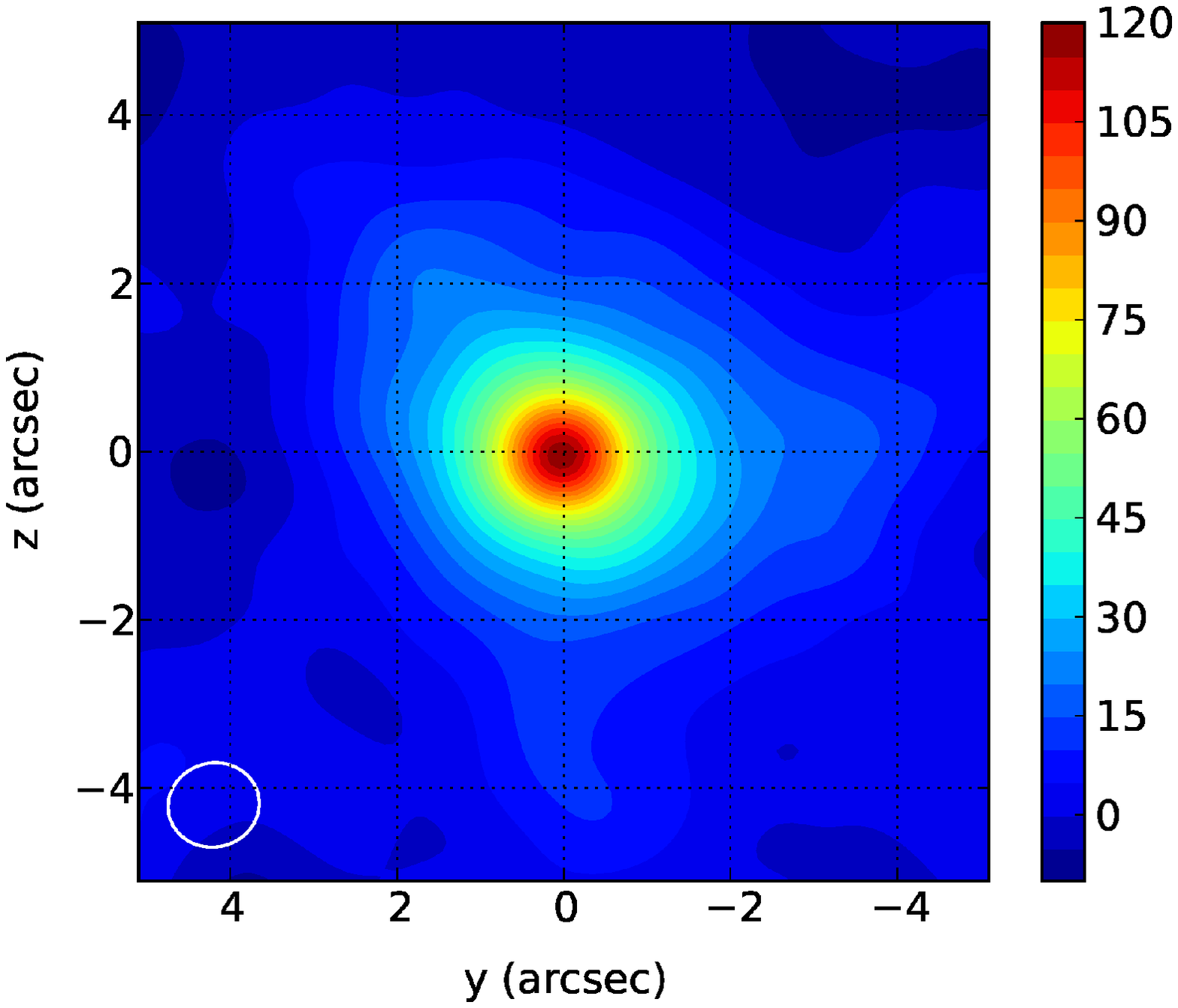}
\includegraphics[scale=0.36]{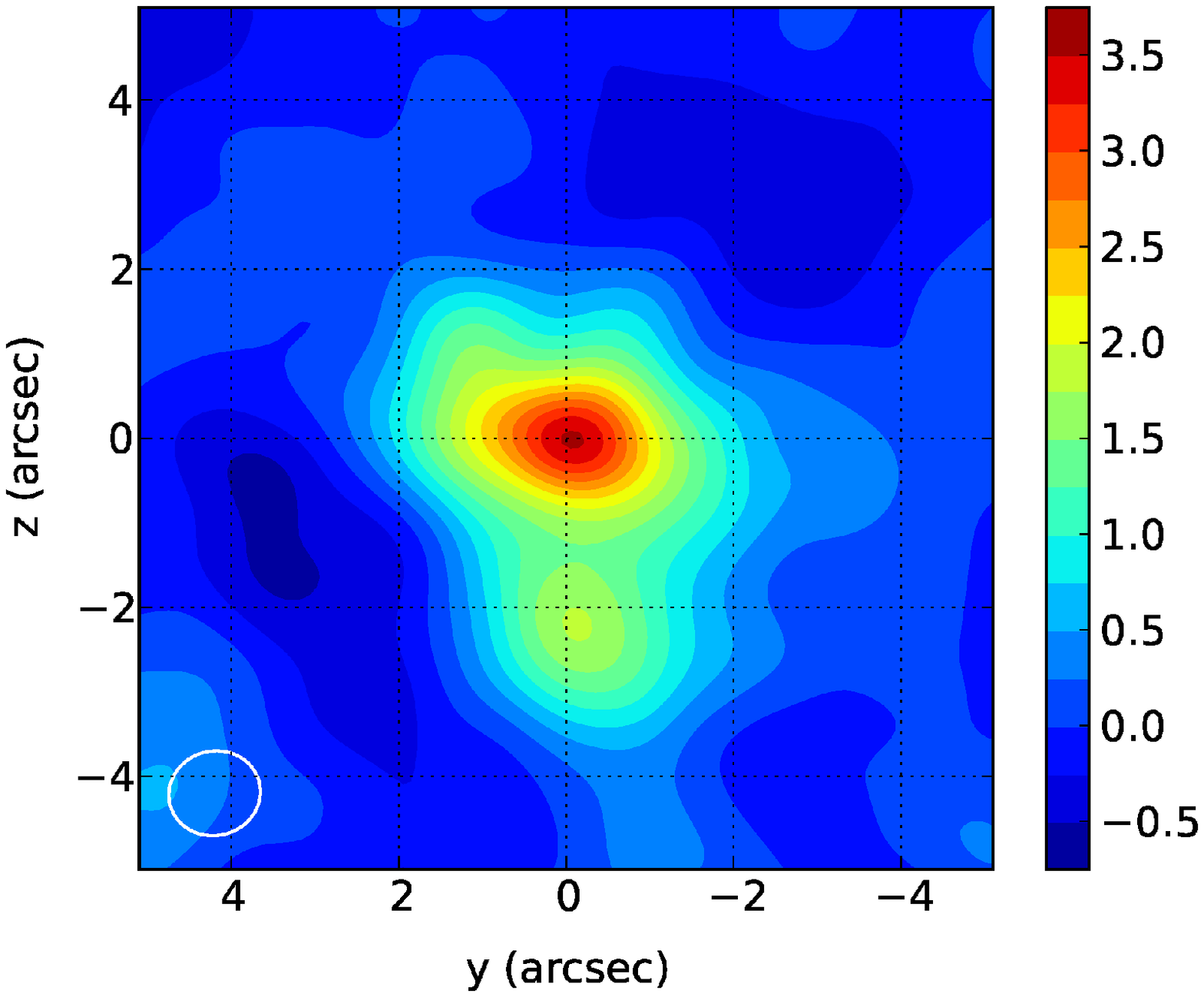}
\caption{ALMA sky maps of the central region for Doppler velocities respectively larger (left) and smaller (right) than
 $-14$ km\,s$^{-1}$. The color scales are in units of Jy km\,s$^{-1}$\,arcsec$^{-2}$.}\label{Fig5}
\end{center}
\end{figure}

Figure \ref{Fig4} displays the sky maps of the detected fluxes over the whole fields of view. In the ACA data, a slight offset of 
the central star with respect to the origin of coordinates, $0.4''$ north and $1.3''$ west, has been corrected for in what follows. 

Figure \ref{Fig5} compares the ALMA sky maps of the central region for Doppler velocities respectively smaller and larger than 
$-14$ km\,s$^{-1}$. The small excess observed in Figure \ref{Fig3} in the [$-$20,$-$14] km\,s$^{-1}$ interval (also present in single dish data) 
corresponds to short distances from the central star with a clear south enhancement at \mbox{$\sim2''$}. Accordingly, in what 
follows, we restrict Doppler velocities to the interval spanning between $-14$ and $18$ km\,s$^{-1}$ for ALMA data. The mean velocity
 of the selected sample of observations is $V_x^{\ast}$=3.2 km\,s$^{-1}$.

\section{Morphology and kinematics of the CSE at short distances ($R<4''$)}
\label{sec:Morphology and kinematics}

\begin{figure}[!ht]
\begin{center}
\includegraphics[scale=0.36]{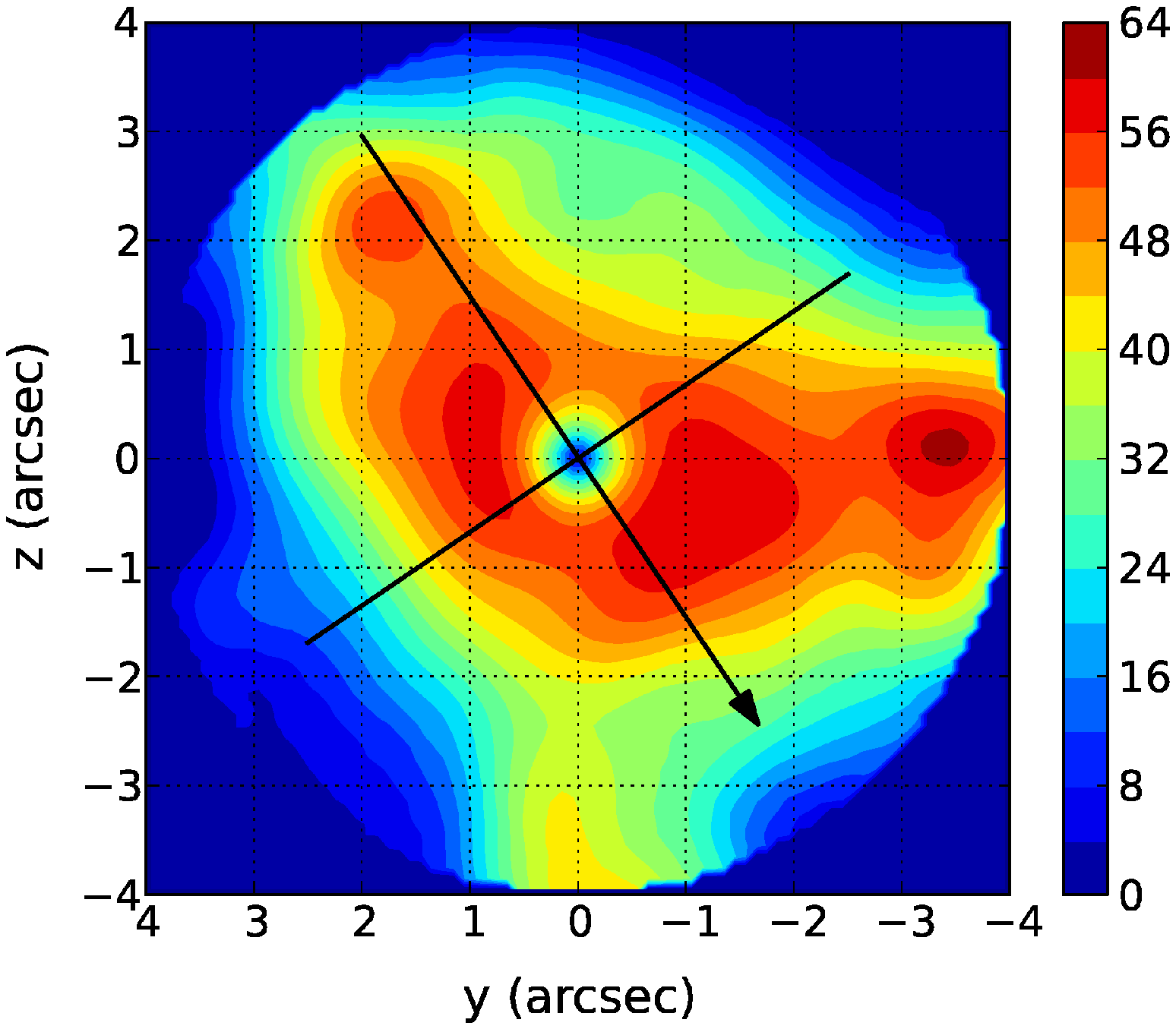}
\includegraphics[scale=0.36]{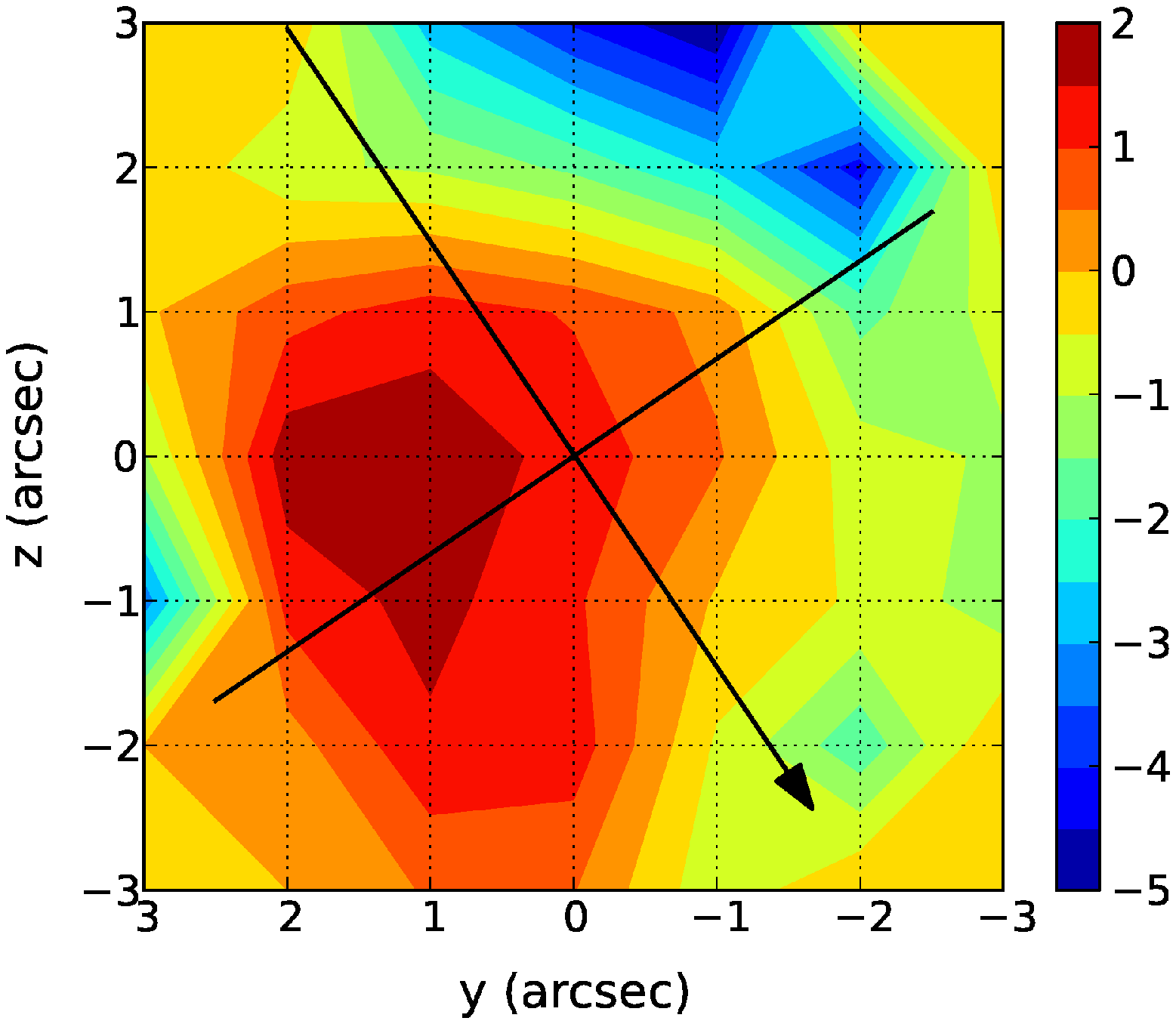}
\caption{Left: sky map of the measured flux (flux density integrated over Doppler velocity from \mbox{$-14$ km\,s$^{-1}$} to 
18 km\,s$^{-1}$) multiplied by $R$ for ALMA data and $R<4''$. Right: sky map of the mean Doppler velocity (averaged between 
$-14$ km\,s$^{-1}$ and 18 km\,s$^{-1}$ and shifted by $V_x^\ast$ to be centred on zero) for ALMA data and $R<3.5''$; large pixel 
sizes ($1''\times1''$) are used in order to keep reasonable signal to noise values in each pixel. The line joining the 
central star to its companion is shown as a black arrow and the other black line is perpendicular to it. The colour scales 
are in Jy km\,s$^{-1}$\,arcsec$^{-1}$ and km\,s$^{-1}$ respectively.}\label{Fig6}
\end{center}
\end{figure}

\begin{figure}[!ht]
\begin{center}
\includegraphics[scale=0.34]{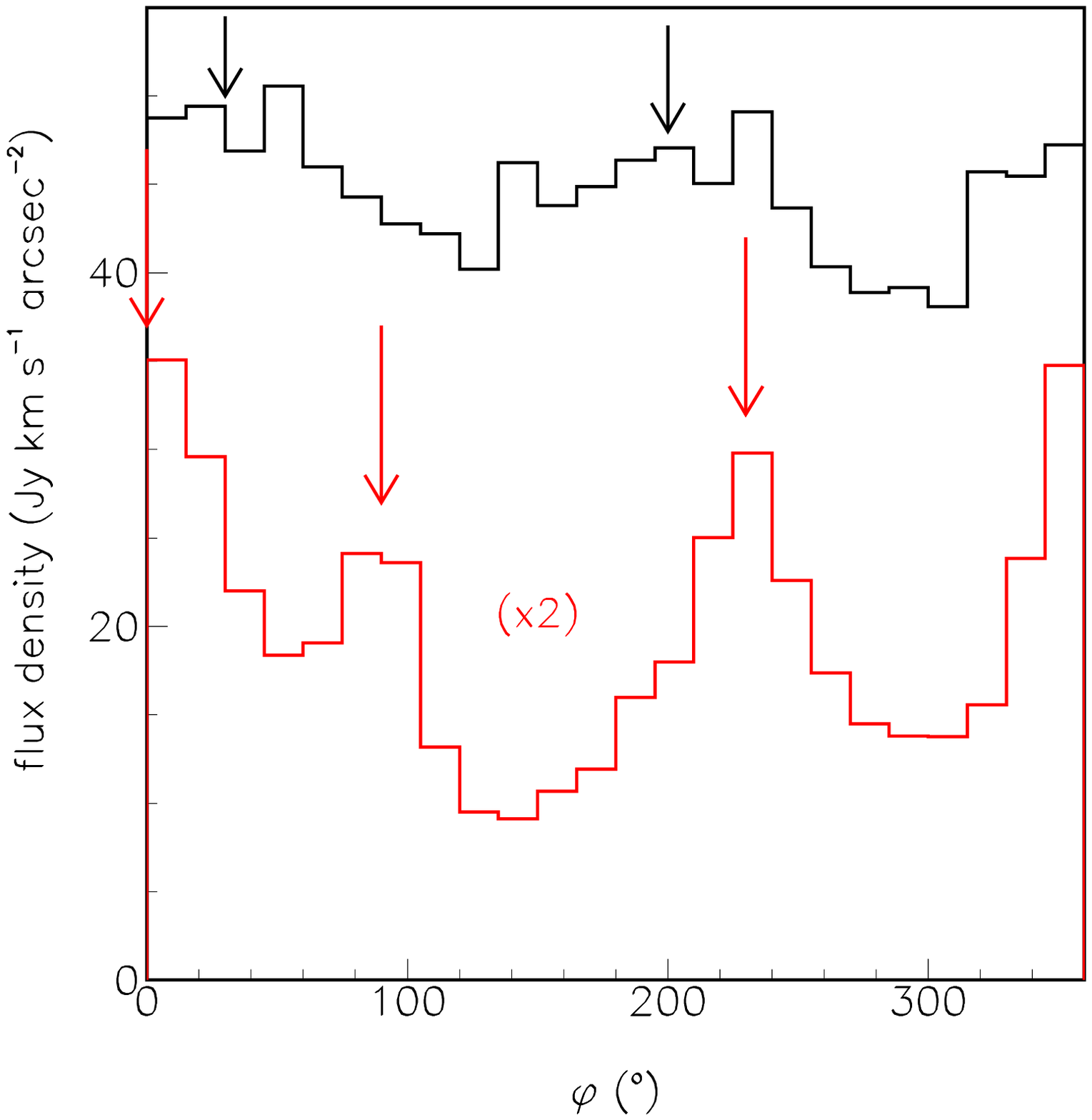}
\includegraphics[scale=0.34]{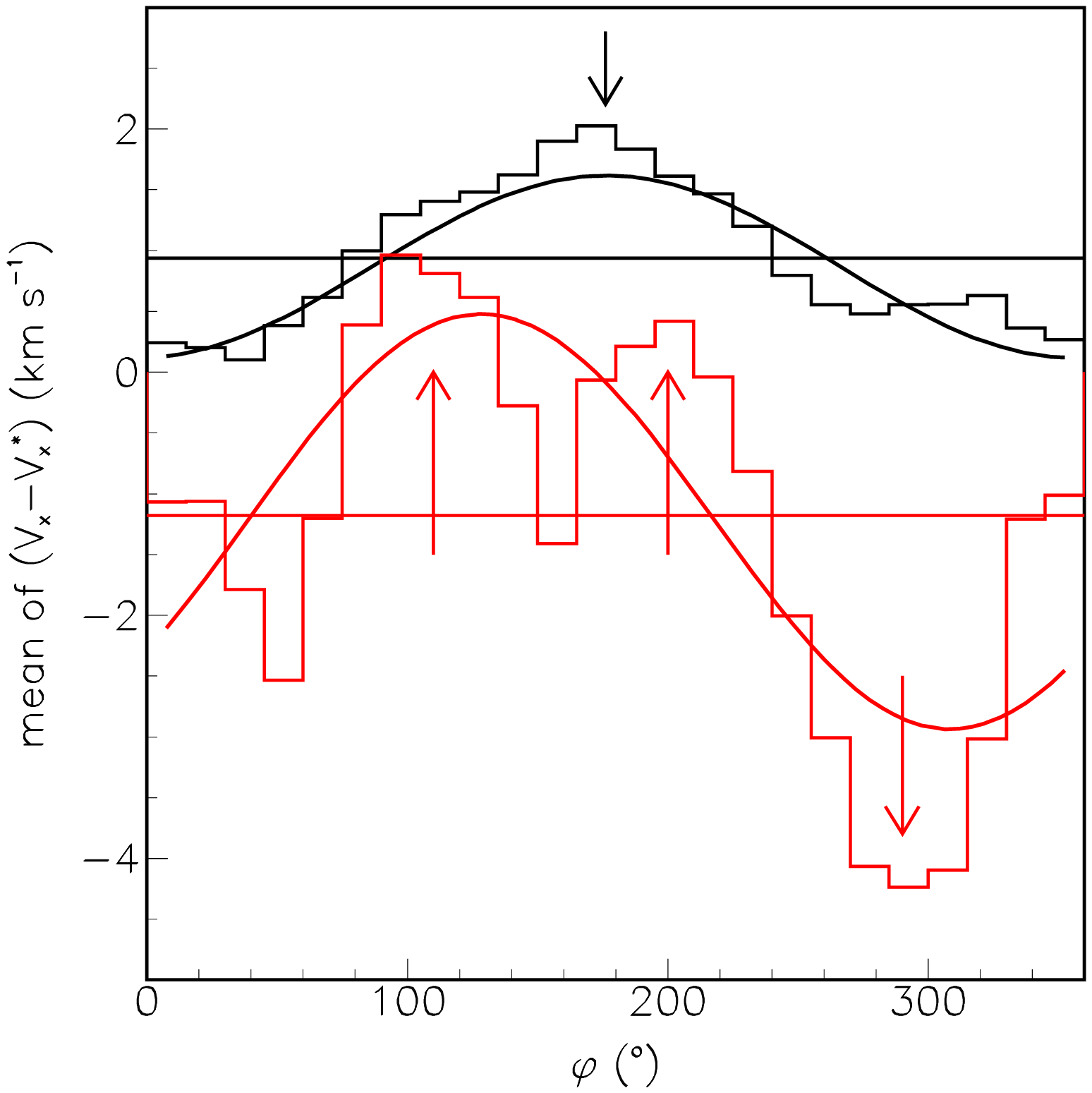}
\caption{Left panel: $\varphi$-distribution of the measured flux for ALMA data and $R<2''$ (black) and $2''<R<4''$ (red, multiplied by 2).
 The arrows indicate approximate locations of the centres of the observed enhancements (30$^\circ$ and 200$^\circ$ in the first case,
 0$^\circ$, 90$^\circ$ and 230$^\circ$ in the second). Right panel: $\varphi$-distribution of the mean Doppler velocity averaged over
 $R<2''$ (black) or over $2''<R<4''$ (red) and over $-14<V_x <18$ km\,s$^{-1}$ (shifted by $V_x^\ast$ to be centred on zero) for 
ALMA data. The arrows indicate approximate locations of the centres of the observed velocity enhancements,
176$^\circ$ in the first case, 110$^\circ$, 200$^\circ$ and 290$^\circ$ in the second). The horizontal lines indicate the mean values 
of $<V_x -V_x^\ast>$ in each of the two $R$ intervals, respectively 0.9 and $-1.2$ km\,s$^{-1}$.}\label{Fig7}
\end{center}
\end{figure}

Figure \ref{Fig6} (left) displays the sky maps of the measured flux (flux density integrated over Doppler velocities from 
$-14$ km\,s$^{-1}$ to 18 km\,s$^{-1}$) multiplied by $R$ for ALMA data and $R<4''$. Multiplying by $R$ would mean a uniform
 distribution for a $1/r^2$ dependence of the emission ($r$ being the space distance from the central star) and allows for 
better revealing inhomogeneities up to large distances. A clear enhancement is seen in the north-east/south-west direction,
 covering a broad angular range, and in the opposite direction, covering a narrower angular range. The broad enhancement is
 centred on $\sim40^\circ$ in the $R<2''$ interval and splits in two components when $R$ increases, respectively centred on
 $\sim0^\circ$ and $\sim90^\circ$ in the $2''<R<4''$ interval. This is also clearly seen in Figure \ref{Fig7} where the measured
 flux has been averaged over $R<2''$ and $2''<R<4''$ separately in $15^\circ$ wide $\varphi$ bins.

Figure \ref{Fig6} (right) shows the sky map of the mean Doppler velocity, (averaged between $-14$ km\,s$^{-1}$ and 18 km\,s$^{-1}$ and 
shifted by $V_x^\ast$ to be centred on zero) for ALMA data and $R<3.5''$; large pixel sizes ($1''\times1''$) are used in order to 
keep reasonable signal to noise values in each pixel. A clear asymmetry is revealed along the direction perpendicular to the line
 joining the central star to its companion, the south-east hemisphere being red-shifted and the north-west hemisphere blue-shifted. 
Again, this is clearly seen in Figure \ref{Fig7} which displays the $\varphi$-distribution of the mean Doppler velocity averaged 
over $R<2''$ and over $2''<R<3.5''$ separately and shifted by $V_x^\ast$ to be centred on zero for ALMA data. Positive velocities
 are observed over a broad angular range in the south-east direction and negative velocities over a narrower angular range in the 
north-west direction. Sine wave fits of the form \mbox{$<V_x -V_ x^\ast>=V_0 +V_ 1 \cos(\varphi-\varphi_0)$} give 
$V_0 =0.87$ km\,s$^{-1}$, $V_1 =0.75$ km\,s$^{-1}$ and $\varphi_0 =176^\circ$ for $R<2''$ and $V_0 =-1.23$ km\,s$^{-1}$,
 $V_1 =1.71$ km\,s$^{-1}$ and $\varphi_ 0 =128^\circ$ for $2''<R<4''$. However, the quality of the fit is poor in the latter case, 
the distribution being split in two maxima centred at $\sim 100^\circ$ and $\sim 190^\circ$ respectively.

%%%%%%%%%%%% Figure 6 %%%%%%%%%%%%%%%%%%%%%%%% 
%%%%%%%%%%%%%%%%%%%%Figure 7%%%%%%%%%%%%%%%%%%%%%%%%%%

\begin{figure}[!ht]
\begin{center}
\includegraphics[scale=0.36]{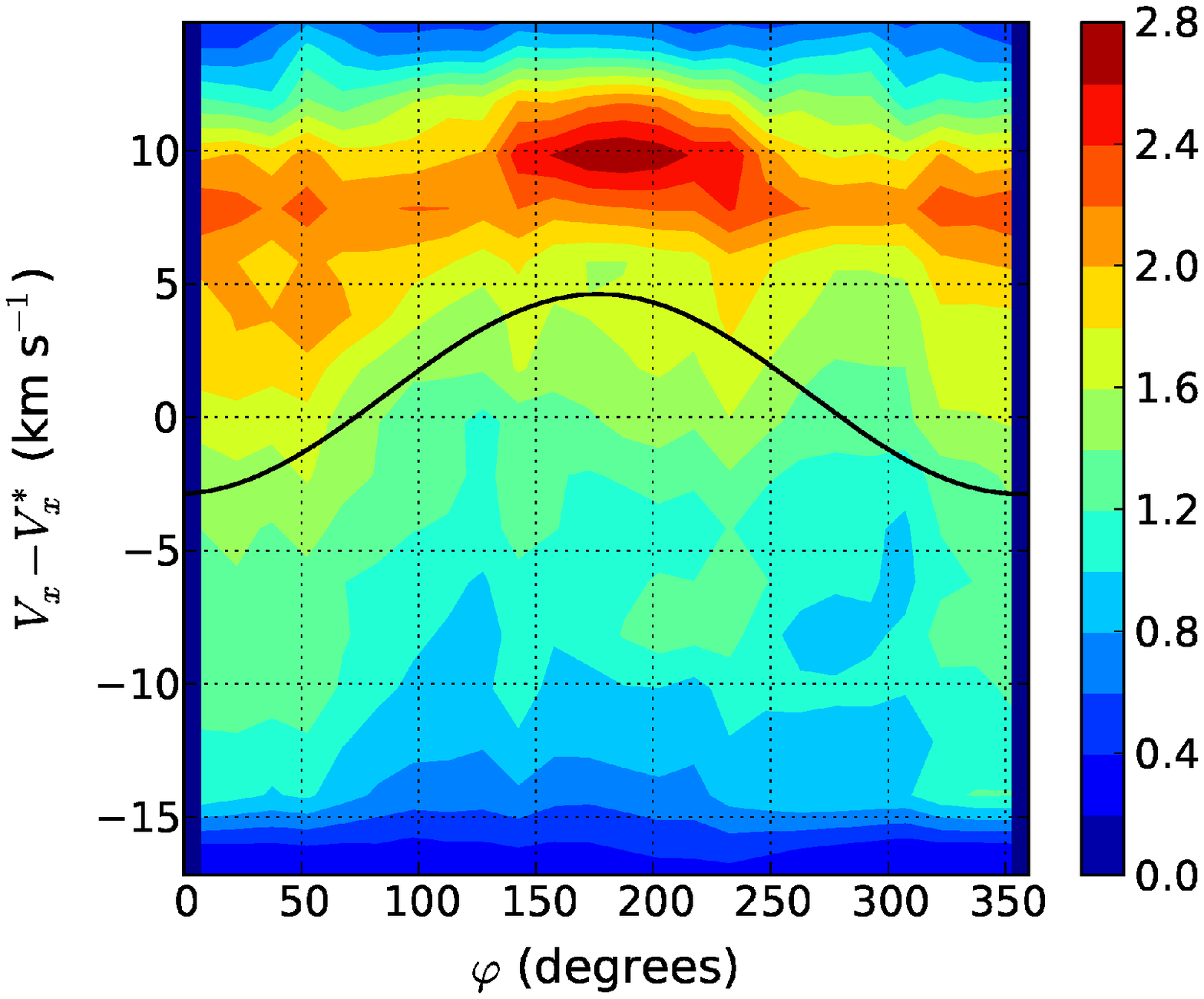}
\includegraphics[scale=0.36]{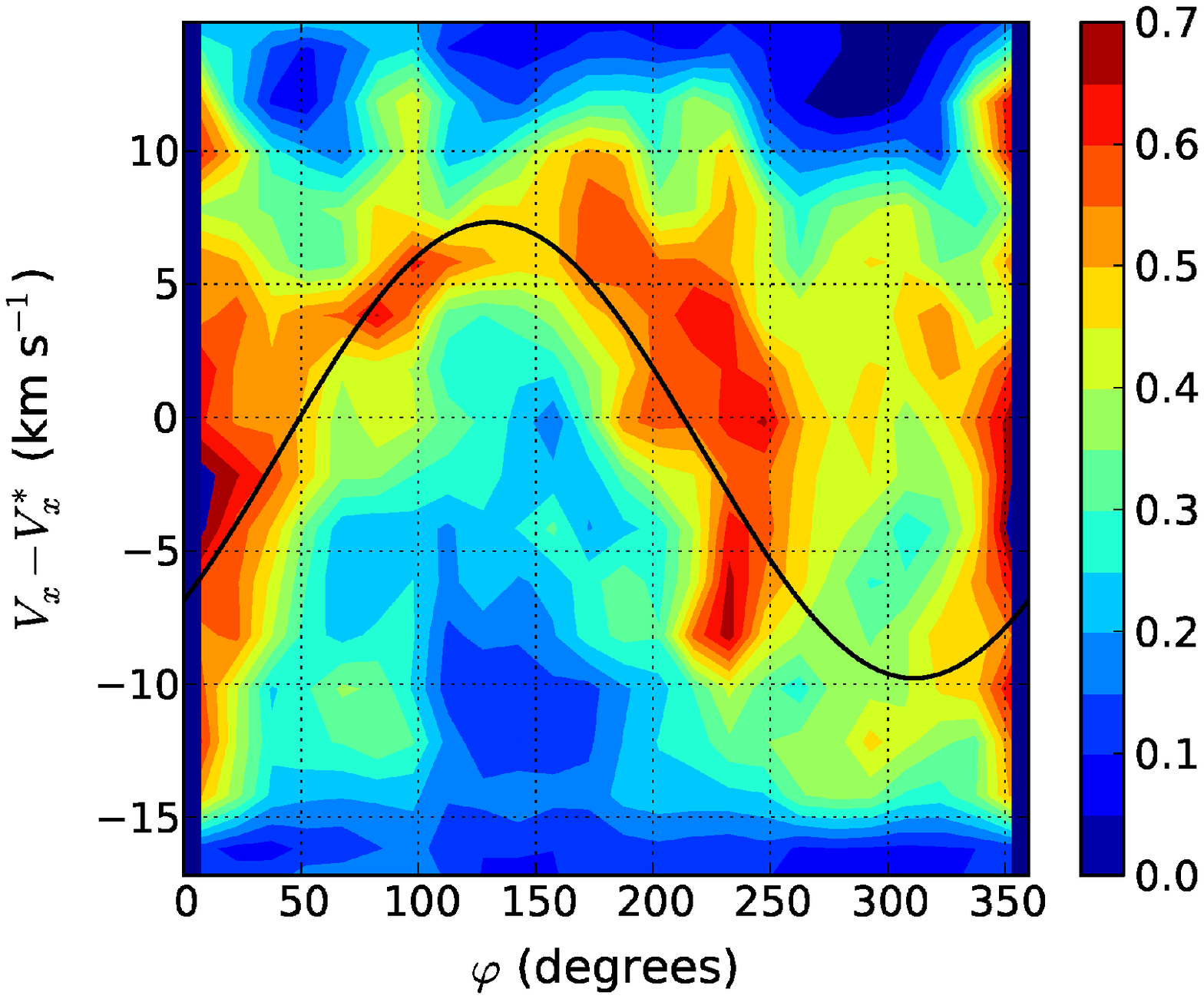}
\caption{Position velocity diagrams averaged over \mbox{$R<2''$} (left panel) and over \mbox{$2''<R<4''$} (right panel). The ordinate is the 
Doppler velocity shifted by $V_x^\ast$ to be centred on zero. The abscissa is the position angle in degrees measured clockwise from west. 
The sine waves are the fits displayed in Figure \ref{Fig7} (right) with amplitudes multiplied by 5.}\label{Fig8}
\end{center}
\end{figure}

Figure \ref{Fig8} shows position-velocity diagrams in the $V_x$ vs $\varphi$ plane for ALMA data averaged over $R<2''$ and over $2''<R<4''$ 
separately. What we call here a position-velocity diagram is different from what is normally meant by this term in the literature, namely
 a plot of the Doppler velocity as a function of a cut in the sky plane containing the source. However, it carries information of a similar 
nature and is particularly relevant when the radial dependence of the Doppler velocity is not too strong. As previously, $V_x$ has been 
shifted by $V_x^\ast$ to be centred on zero. The sine wave fits displayed on Figure \ref{Fig7} (right) are shown in Figure \ref{Fig8} with
 amplitudes multiplied by 5. In the first case, the maxima of emission correspond approximately to the extrema of velocity while in the 
second case they are in approximate quadrature with the extrema of velocity. Qualitatively, the situation can be summarized as follows: 
when scanning across the $R<4''$ interval, one evolves from a morphology elongated along the east/west direction with a velocity gradient 
directed along this same direction to a morphology elongated along the north-east/south-west direction (with a broad enhancement south-west 
and a narrow enhancement north-east) and a velocity gradient directed along the perpendicular to it. Moreover, the $R<2''$ region is 
red-shifted with respect to the $2''<R<4''$ region by $\sim2.1$ km\,s$^{-1}$.

%%%%%%%%%%%%%%%% Figure 8 %%%%%%%%%%%%%%%%%%%%%%%%%%%%%%%

\begin{figure}[!ht]
\begin{center}
\includegraphics[scale=0.33]{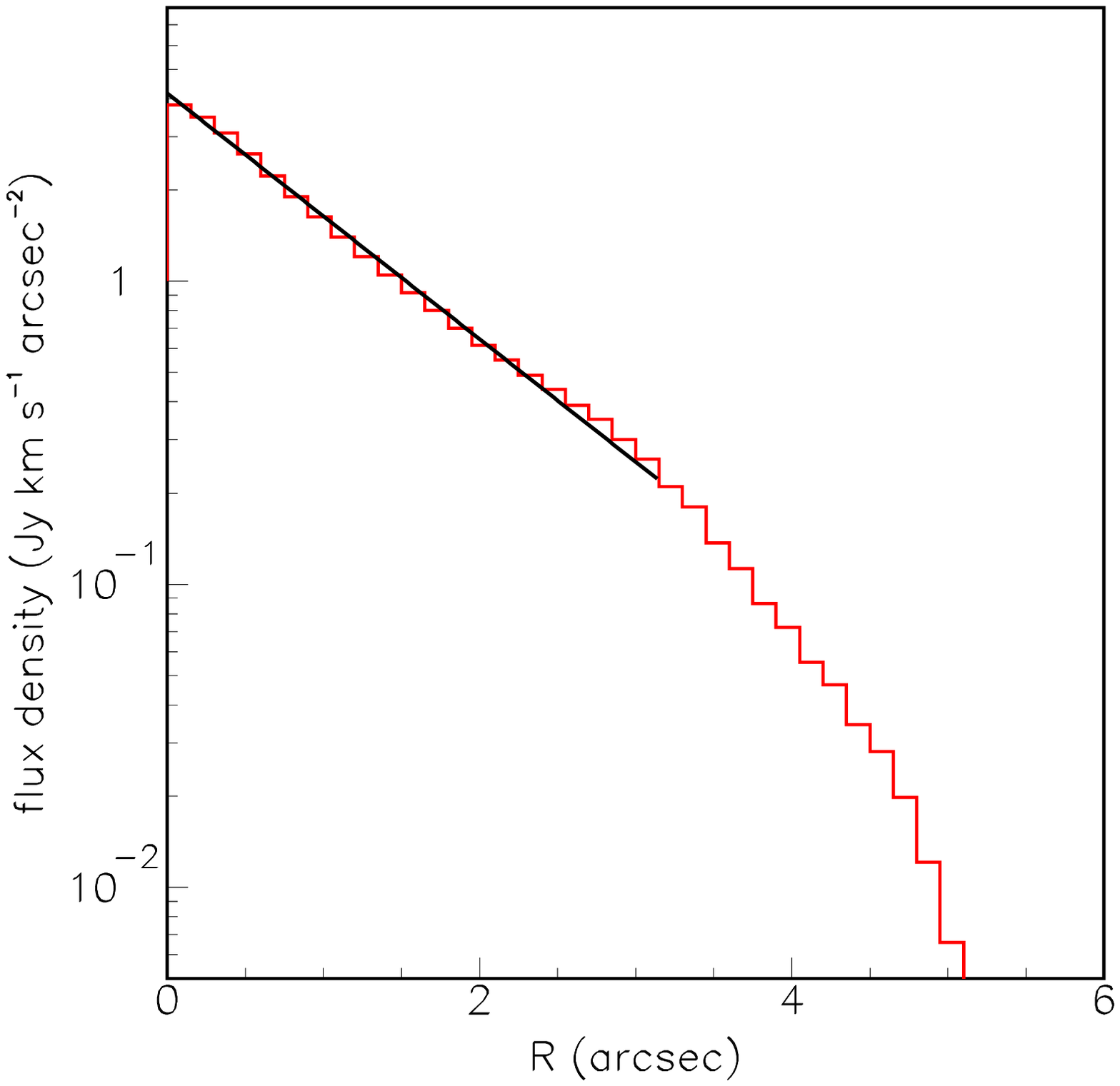}
\includegraphics[scale=0.33]{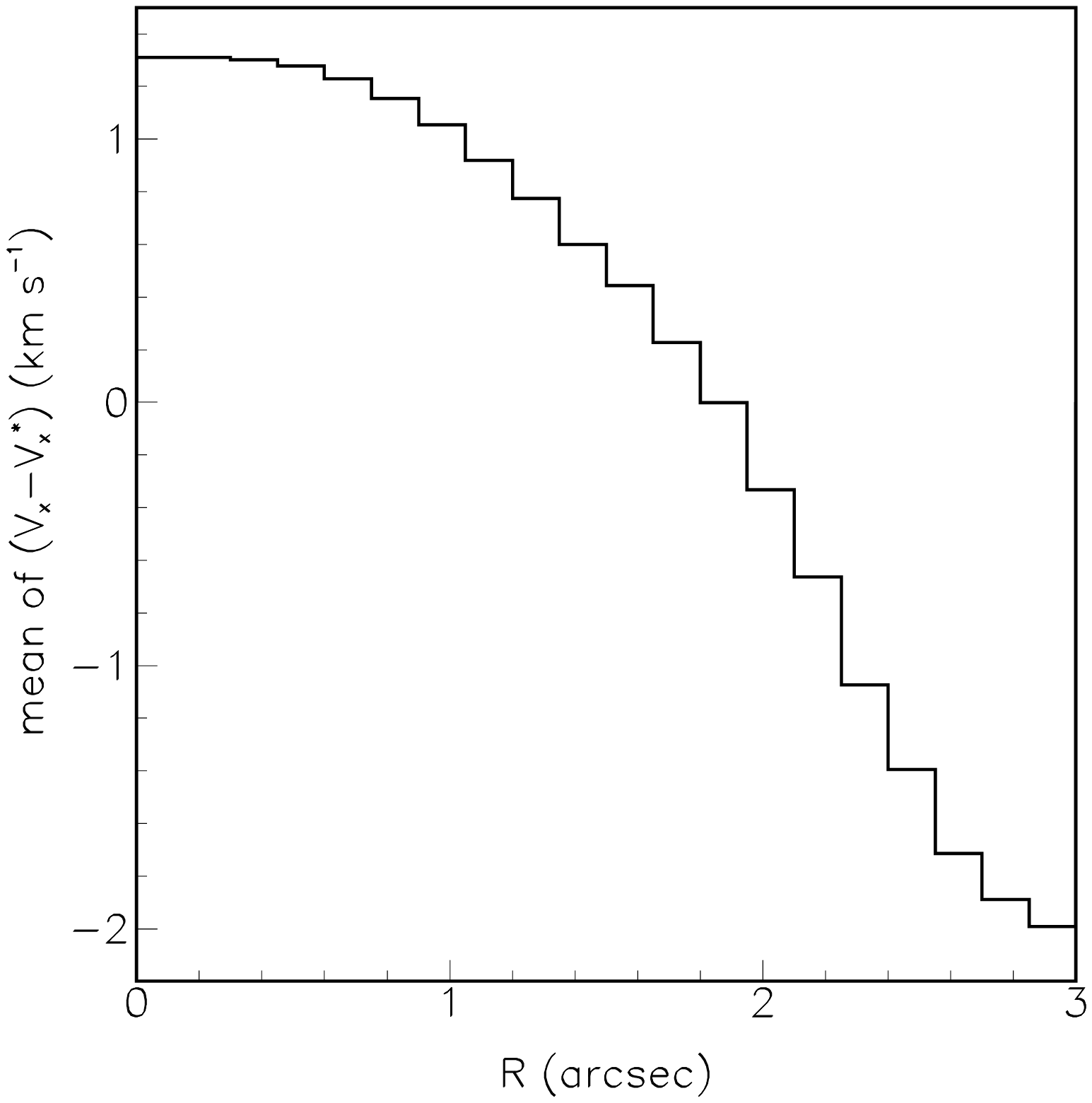}
\caption{Left: $R$-distribution of the flux averaged over $\varphi$ for ALMA data; the line is an exponential fit of the form $4.18e^{-R/1.08''}$.
 Right: dependence on $R$ of the mean Doppler velocity, shifted by $V_x^\ast$ to be centred on zero.}\label{Fig9}
\end{center}
\end{figure}

Figure \ref{Fig9} (left) shows the $R$ distribution of the flux averaged over $\varphi$ for ALMA data. An exponential fit of the form
 $A_0 e^{-R/R_0}$ for $R<3''$ gives $A_0=4.18$ Jy km\,s$^{-1}$\,arcsec$^{-2}$ and $R_0=1.08''$. The inhomogeneities apparent in the left
 panel of Figure \ref{Fig6}, where the flux is multiplied by $R$, are too small to be visible. The drop beyond $4''$ is the effect of 
the short spacing problem. Also shown in Figure \ref{Fig9} (right) is the dependence on $R$ of $<V_x -V_x^\ast>$. It decreases rapidly 
when $R$ exceeds $\sim1.5''$.

%%%%%%%%%%%%%%%% Figure 9 %%%%%%%%%%%%%%%%%%%%%%%%%%%%%%%

%%%%%%%%%%%%%%%% Figure 10 %%%%%%%%%%%%%%%%%%%%%%%%%%%%%%%
\begin{figure}[!ht]
\begin{center}
%trim option's parameter order: left bottom right top
\includegraphics[scale=0.33]{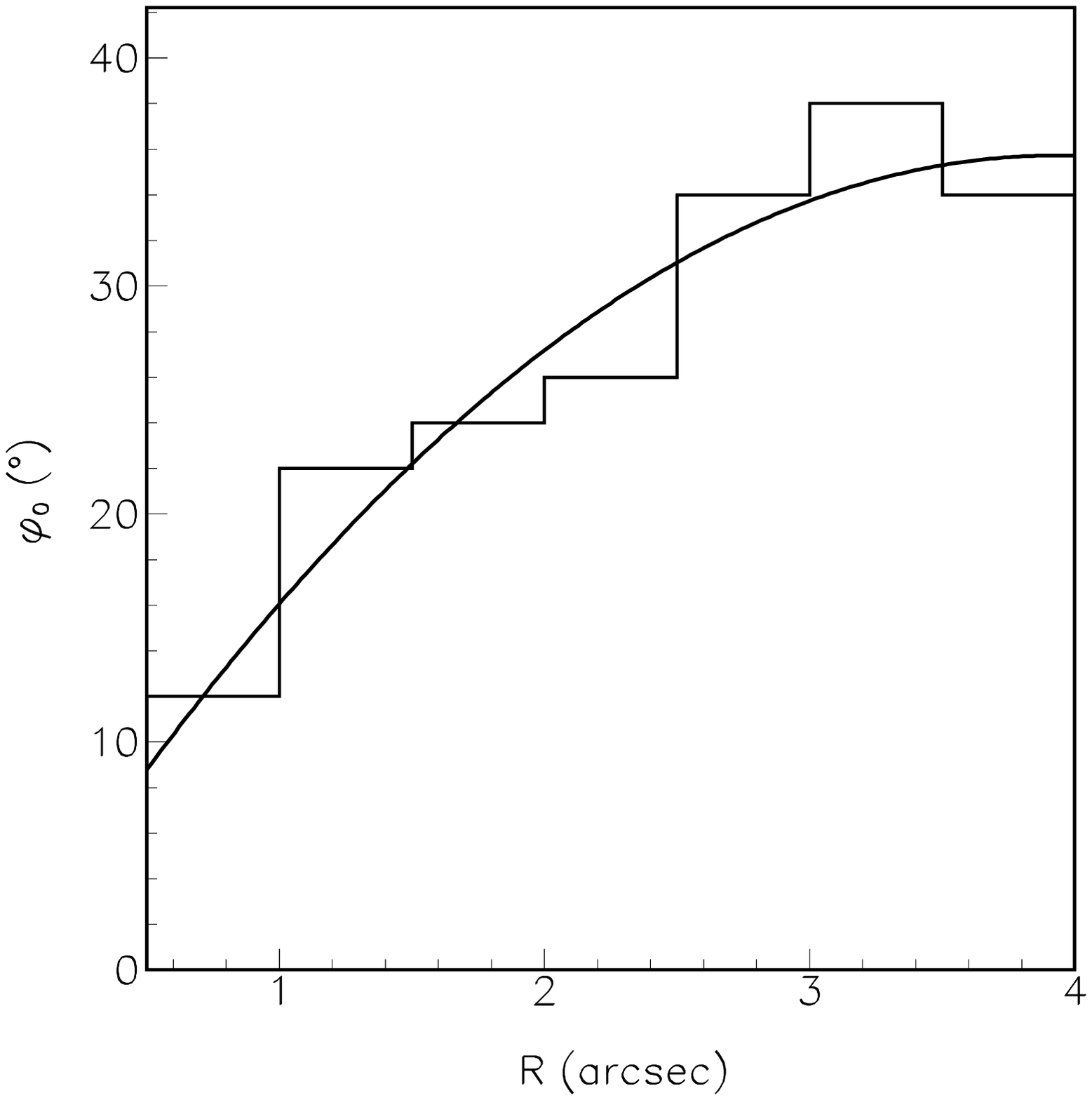}
\includegraphics[scale=0.33]{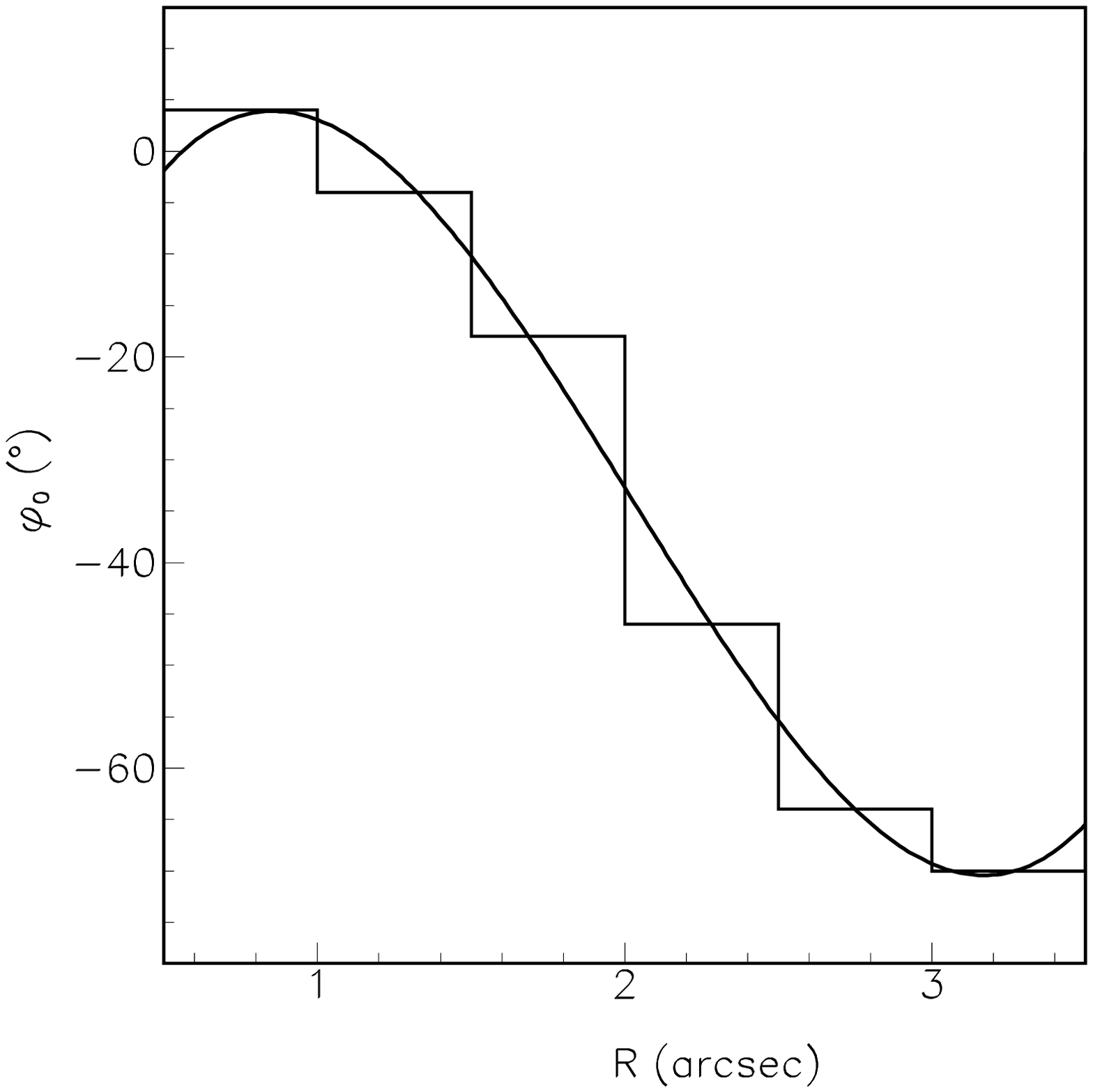}
\caption{Left: $R$-distribution of the direction of the flux maximal elongation. Right: $R$-distribution of the direction of maximal 
velocity asymmetry. The curves are polynomial fits to guide the eye.}\label{Fig10}
\end{center}
\end{figure}

Figure \ref{Fig10} illustrates the evolution of the direction of elongation of the emission and of the wind asymmetry when $R$ increases.
 Here, directions of elongation are defined as the values of angle $\varphi_0$ that maximize the mean values of \mbox{$\cos^2(\varphi-\varphi_0)$} 
and \mbox{$(V_x-V_x^\ast)\cos(\varphi-\varphi_0)$} respectively, weighted by the flux in each bin of $R$ ($0.5''$ wide). Here, $V_x^\ast$ 
has been corrected to be the mean value of $V_x$ in the relevant $R$-bin. They show very clearly the evolution of the flux maximal elongation 
from $\sim10^\circ$ above west to $\sim 10^\circ$ below north-west and that of the maximal velocity asymmetry from west to $\sim20^\circ$ above north-west.

\section{Morphology and kinematics of the CSE at large distances}
\label{sec:large distances}

\begin{figure*}[!ht]
\begin{center}
%trim option's parameter order: left bottom right top
\includegraphics[trim = 0cm .8cm 0cm 0cm, clip, scale=0.36]{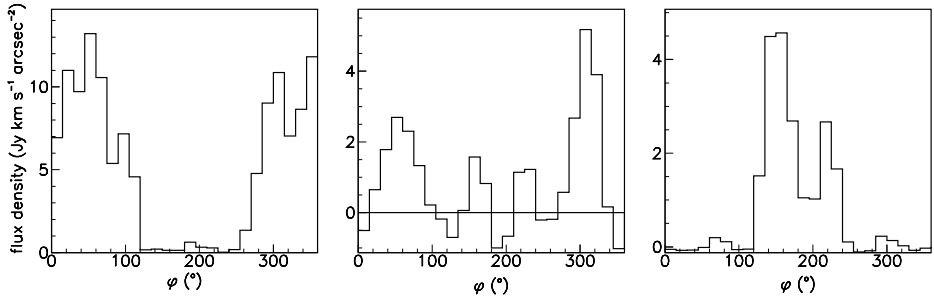}
\caption{Distribution on $\varphi$ of the flux measured with the compact ACA array, averaged over \mbox{$5''<R<10''$ (left)}, \mbox{$10''<R<15''$ (centre)} and \mbox{$15''<R<20''$ (right)}.}\label{Fig11}
\end{center}
\end{figure*}

\begin{figure}[!ht]
\begin{center}
\includegraphics[trim = 0cm -3.6cm 0cm 0cm, clip, scale=0.205]{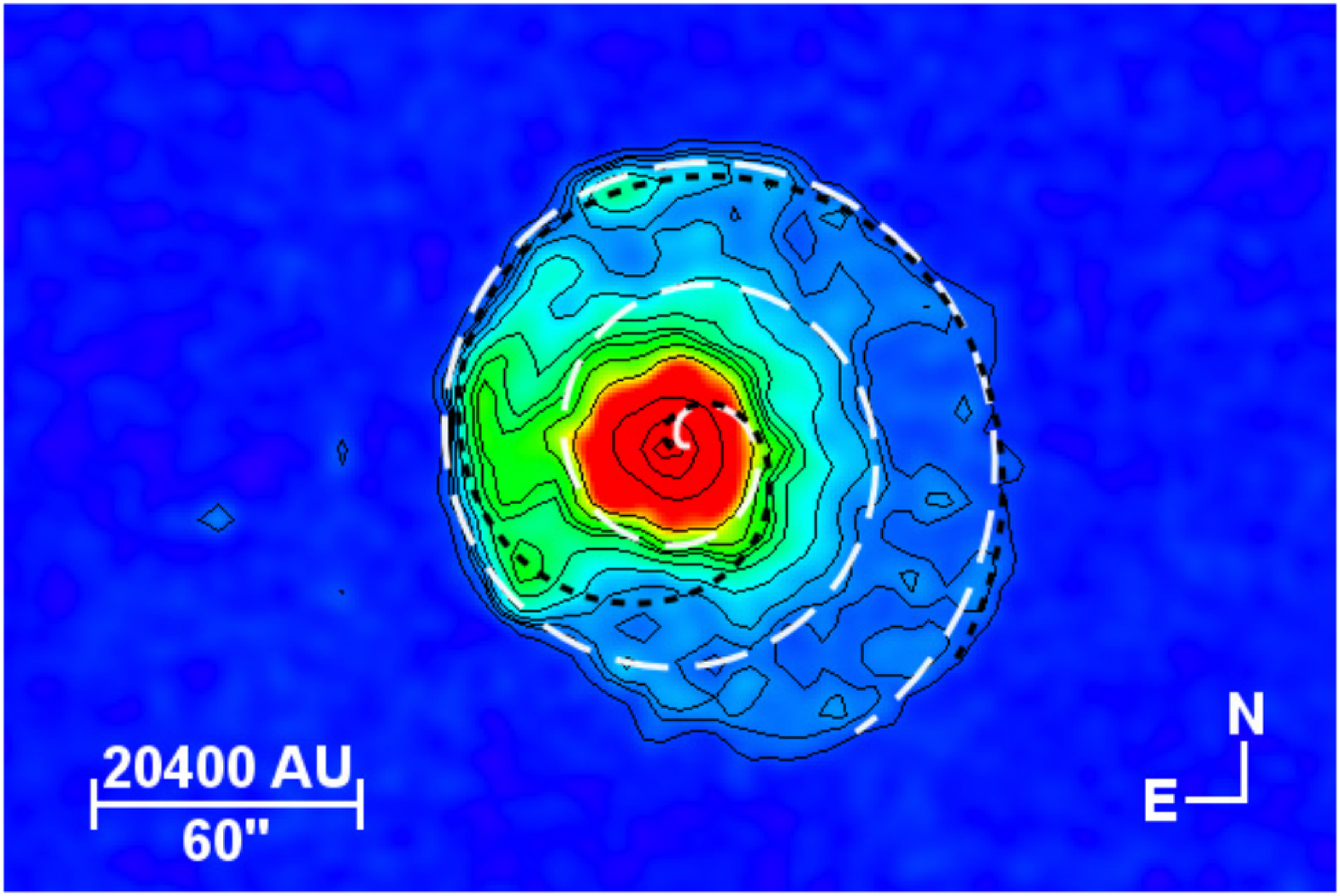}
\includegraphics[scale=0.33]{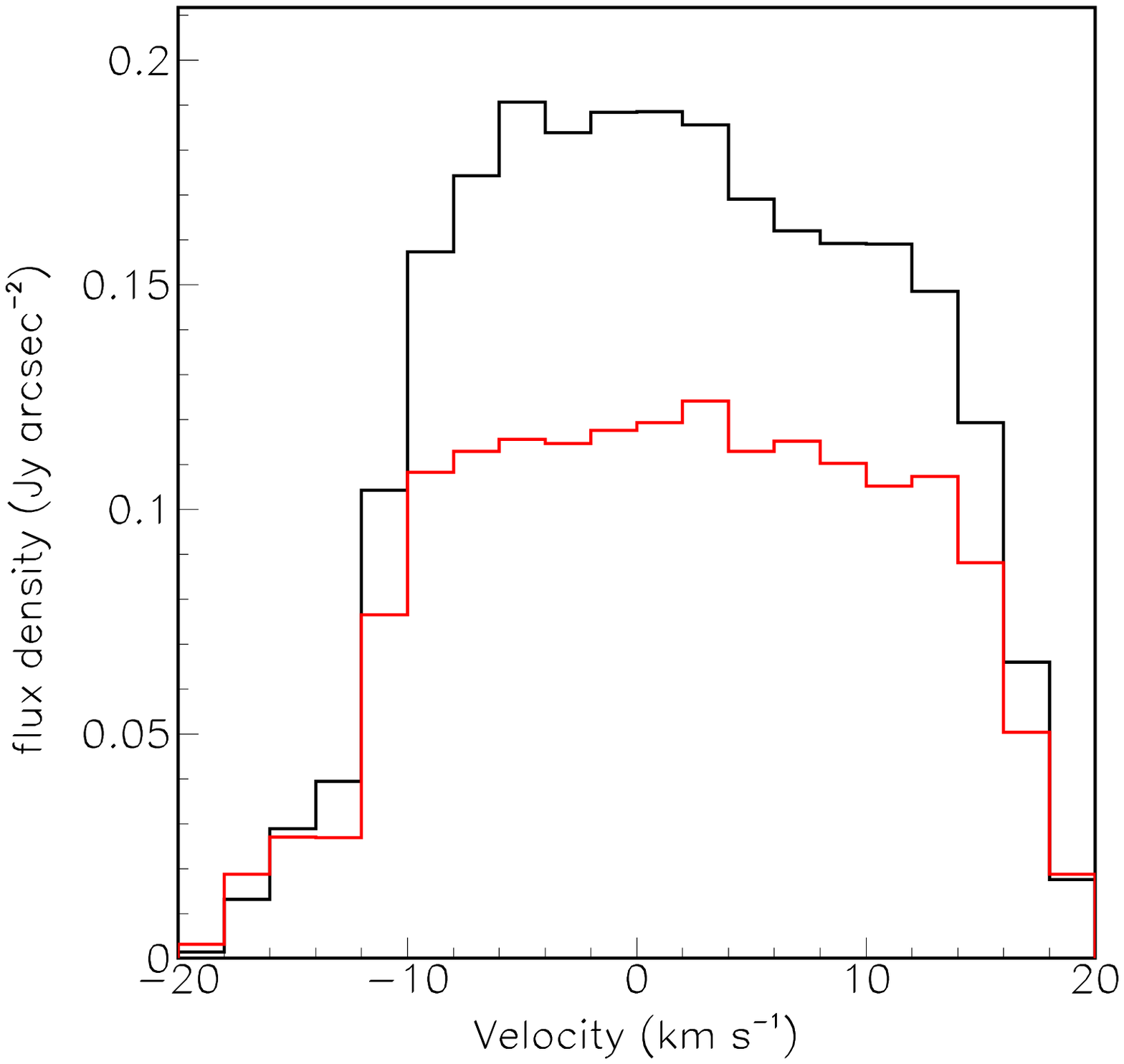}
\caption{Left: large scale environment of W Aql observed by Mayer et al. (2013) at 70 $\mu$m from Herschel/PACS. 
Right: velocity spectra averaged over the enhancements displayed in Figure \ref{Fig11}: \mbox{$5''<R<10''$} and 
$|\varphi|<\pi/2$ (black) and \mbox{$15''<R<20''$} and $|\varphi–\pi|<\pi/3$ (red).}\label{Fig12}
\end{center}
\end{figure}

The dusty CSE of W Aql is known to extend up to large distances, typically $50''$ east and $70''$ west, contained within an envelope described as 
a spiral by Mayer et al. (2013). From Mamon et al. (1988) we obtain an effective UV dissociation distance for CO molecules 
$r_{1/2}\sim$ 1.5 10$^{17}$ cm, meaning \mbox{$\sim10^4$ AU} or $\sim25''$. The $\varphi$-distribution of the flux measured 
with the compact ACA array is displayed in \mbox{Figure \ref{Fig11}} for three $R$ intervals, $5''$ to $10''$, $10''$ to 
$15''$ and $15''$ to $20''$. They show clear enhancements in the western half-plane for $5''<R<10''$, joining smoothly with 
the enhancements observed in \mbox{Figure \ref{Fig6} (left)} for $R<4''$, and at $\pm60^\circ$ from east for $15''<R<20''$, 
joining smoothly with the eastern enhancement observed by Mayer et al. (2013) for $R>20''$ (Figure \ref{Fig12} left). 

%%%%%%%%%%%%%%%%%%%%%%% Fig 11 %%%%%%%%%%%%%%%%
%%%%%%%%%%% + Figure 12%%%%%%%%%%%%%%%%%%%%%%%%

Figure \ref{Fig12} (right) displays the Doppler velocity distributions measured over each of these enhancements; they 
do not show any particular feature. Their mean velocities are 1.8 km\,s$^{-1}$ and 2.0 km\,s$^{-1}$ compared with 
$V_x^\ast$=3.2 km\,s$^{-1}$ obtained at small distances from the central star for ALMA data. The present data do not
 allow for drawing a more detailed picture of the CSE at large distances from the star.

\section{Discussion and summary}
\label{sec:Discussion}

The analyses of the preceding sections show clearly that longer observation times, preferably including lines associated
 with several CO excitations, are needed in order to reach a reliable and detailed understanding of the morphology and
 kinematics of the CSE of W Aql. Yet, some qualitative but important results have been obtained that bring significant 
new information at the same time as they raise new questions.

The important slope of the line profile in the central region, of the order of 2\% per km\,s$^{-1}$, is a surprising observation 
as is the rapid decrease of $<V_x >$ with $R$ when $R$ exceeds $\sim2''$. This may be an effect of optical thickness, 
which is expected to affect more strongly small values of $R$, therefore to increase the measured value of the apparent
 mean Doppler velocity at small values of $R$. If such is the case, it raises the question of the importance of optical
 thickness on the interpretation of the results presented in the preceding sections. To a first approximation, one would 
expect the $\varphi$-dependence of the measured quantities at fixed values of $R$ not to be much affected.

The measurement of the star velocity in the local standard of rest (LSR) frame is made difficult by the variability of the
 star and its interaction with its companion. Mayer et al. (2013) have presented a detailed analysis of available data and 
have concluded that the space velocity of W Aql is \mbox{$V_{LSR}=21.6\pm4.1$ km\,s$^{-1}$} with a position angle of 
38$^\circ$ and an inclination of $-17^\circ$ with respect to the line of sight. Our value of $V_x^\ast$, 3.2 km\,s$^{-1}$,
 would therefore suggest $-25.5+3.2=-22.3$ km\,s$^{-1}$ compared with $-21.6 \cos(17^\circ)=-20.8$ km\,s$^{-1}$ and with
 $-23.0$ km\,s$^{-1}$ used by Danilovich et al. (2014), namely a good agreement given the measurement uncertainties, 
in particular the dependence on $R$, probably the effect of optical thickness.

The large mass loss rate of W Aql has significantly shaped its CSE away from spherical. The complex structure of the wind 
velocity field suggests that different regimes have dominated over time. The approximate coincidence between the projected
 position of the companion and the direction of the dominant winds may be accidental: to decide on this important issue 
requires more detailed information than available here.

A velocity gradient along a direction projecting at position angle $\varphi_0$ in the plane of the sky may be interpreted, 
among others, as an expansion along an axis projecting at position angle $\varphi_0$ or as a rotation about an axis projecting
 at position angle $\varphi_0+\pi/2$. In both cases the axis should have a significant inclination with respect to the line
 of sight for the gradient to be observed. At short distances ($R<4''$) the present data do not allow
for choosing between these two interpretations, or possibly suggesting a different one.

When new ALMA data will become available, the understanding of the properties of the CSE of \mbox{W Aql} will require the 
use of a model taking due account of radiative transfer in order to properly describe the effect of absorption at short
 projected distances from the central star, as the measured line profile, and its dependence over $R$, suggest significant
 optical thickness in this central region. For the model parameters, in particular the temperature, to be well constrained
 by observations, it will be necessary to observe emission from at least two different rotational lines of the CO molecule.

The presence of spikes at the extremities of the Doppler velocity distribution, for $|V_x|$ between $\sim14$ and $\sim 20$ km\,s$^{-1}$, 
is not understood. While the blue-shifted spike is enhanced at small distances south of the central star (Figure \ref{Fig5}), 
no similar enhancement can be revealed for the red-shifted spike.

Finally, the observation of \mbox{CO(3-2)} emission at large distances from the central star using the compact ACA array is 
consistent with the results obtained earlier from dust emission in the far infrared but the present data are not sufficient 
to reveal details such as a possible spiral structure as observed by Mayer et al. (2013).

\section*{ACKNOWLEDGEMENTS}
 We are indebted and very grateful to the ALMA partnership, who are making their data available to the public after a one 
year period of exclusive property, an initiative that means invaluable support and encouragement for Vietnamese astrophysics.
 We particularly acknowledge friendly and patient support from the staff of the ALMA Helpdesk. We express our deep gratitude 
to Professors Nguyen Quang Rieu and Thibaut Le Bertre for having introduced us to radio astronomy and to the physics of 
evolved stars. Financial support is acknowledged from the Vietnam National Satellite Centre (VNSC/VAST), the NAFOSTED funding
 agency, the World Laboratory, the Odon Vallet Foundation and the Rencontres du Viet Nam.

% UNCOMMENT THE LINES BELOW IF YOU WISH TO USE BIBTEX
%\bibliographystyle{apj}
%\bibliography{yourbibfile}
%\begin{thebibliography}{}

\label{lastpage}

\end{document}